\documentclass[12pt]{article}
\usepackage{latexsym}
\usepackage{graphicx}
\newcommand{\beq}{\begin{equation}}
\newcommand{\eeq}[1]{\label{#1}\end{equation}}
\newcommand{\bea}{\begin{eqnarray}}
\newcommand{\eea}[1]{\label{#1}\end{eqnarray}}

\begin{document}
\begin{flushright}

\hfill\ \ \ {MCTP-01-60}\ \ \

{hep-th/0111237}\\
\end{flushright}




\vspace*{0.88truein}

\centerline{\bf THE WORLD IN ELEVEN DIMENSIONS: }
\vspace{0.15truein}
\centerline{\bf A TRIBUTE TO OSKAR KLEIN\footnote{ Oskar Klein Professorship
Inaugural Lecture, University of Michigan, 16 March 2001.  }}
\vspace*{0.37truein} \centerline{\footnotesize M.  J.
DUFF\footnote{mduff@umich.edu.  Research supported in part by DOE
Grant DE-FG02-95ER40899.}} \vspace*{0.015truein}
\centerline{\footnotesize\it Michigan Center for Theoretical Physics}
\baselineskip=10pt
\centerline{\footnotesize\it Randall Laboratory, Department of Physics,
University of Michigan}
\centerline{\footnotesize\it Ann Arbor, MI 48109--1120, USA}
\bigskip
\vspace*{0.21truein}
\abstract{Current attempts to find a unified theory that would reconcile
Einstein's General Relativity and Quantum Mechanics, and explain all
known physical phenomena, invoke the Kaluza-Klein idea of extra
spacetime dimensions.  The best candidate is M-theory, which lives in
eleven dimensions, the maximum allowed by supersymmetry of the
elementary particles. We give a non-technical account.

An Appendix provides an updated version of
Edwin A.  Abbott's 1884 satire {\it Flatland: A Romance of Many
Dimensions}.  Entitled {\it Flatland, Modulo 8}, it describes the
adventures of a superstring theorist, A.  Square, who inhabits a ten-
dimensional world and is initially reluctant to accept the existence
of an eleventh dimension.}

\bigskip\bigskip\bigskip\bigskip\bigskip
{{\it ``Returning to my Ann Arbor attempts, I became immediately
very eager to see how far the mentioned analogy reached, first trying
to find out whether the Maxwell equations for the electromagnetic
field together with Einstein's gravitational equations would fit into
a formalism of five-dimensional Riemann geometry.''}}{}{}

\medskip
Oskar Klein, ``From My Life in Physics''
\bigskip


\setcounter{footnote}{0}
\renewcommand{\thefootnote}{\alph{footnote}}

\vspace*{-0.5pt}
\newpage
\section{Physics of the new millennium}
\label{Physics}
\noindent At the end of the last millennium, it became
fashionable for certain pundits to declare the {\it End of Science},
on the grounds that all the most important scientific discoveries had
already been made.  To a physicist, this seems absurd because the two
main pillars of twentieth century physics, namely Einstein's General Theory
of Relativity and Quantum Mechanics, are mutually incompatible.  At the
microscopic level, general relativity fails to comply with the quantum rules
which govern the behavior of the elementary particles; while on the
macroscopic scale, the black holes of Einstein's theory are
threatening the very foundations of quantum mechanics.  Something big has to
give.  This augurs less the bleak future of diminishing returns
predicted by the millennial Jeremiahs and more another scientific
revolution.

\begin{figure}[t]\centering\includegraphics[scale=0.5]{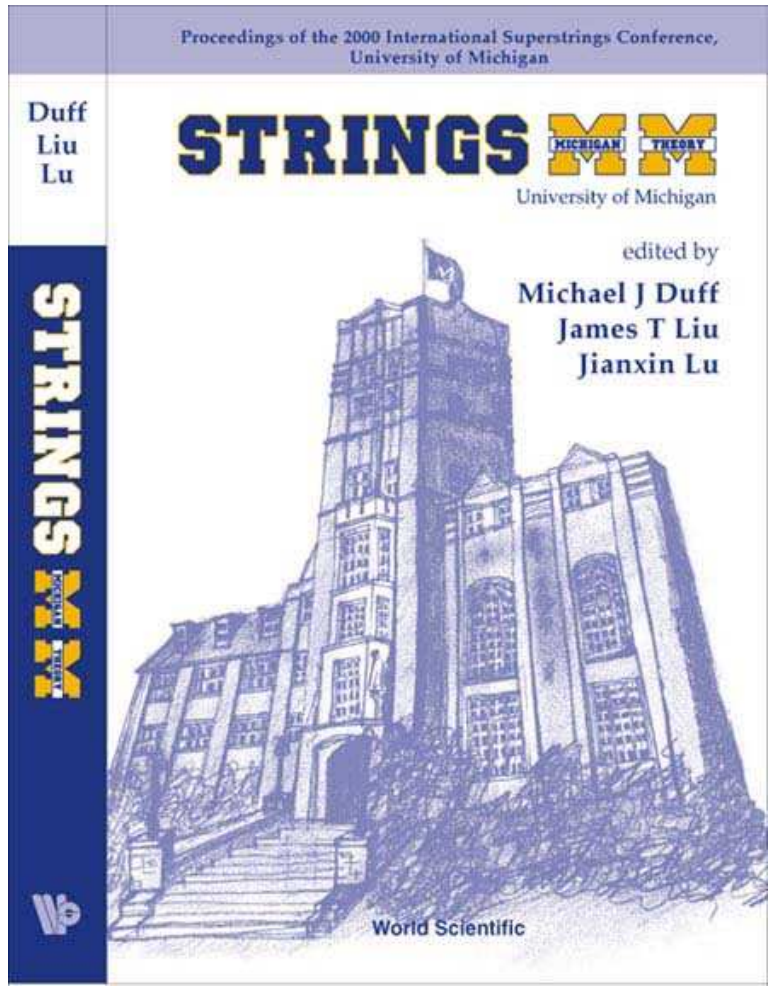}
\caption{The Strings 2000 Conference.}\label{Strings}\end{figure}

Many physicists believe that this revolution is already under way
with the theory of {\it Superstrings}. As their name suggests,
superstrings are one-dimensional string-like objects
\cite{Greene}. Just like violin strings, these relativistic
strings can vibrate and each mode of vibration corresponds to a
different elementary particle. One strange feature of superstrings
is that they live in a universe with nine space dimensions and one
time dimension. Since the world around us seems to have only three
space dimensions, the extra six would have to be curled up to an
unobservably small size (or else rendered invisible in some other
way) if the theory is to be at all realistic. This idea of extra
dimensions will be an important theme in this lecture. Many of you
in this audience will be familiar with superstring theory
following the millennial edition of the annual international
superstrings conference ``Strings 2000'' hosted last year here at
the University of Michigan \cite{Proceedings}.  See Figure
\ref{Strings}.

\begin{figure}\centering\includegraphics[scale=0.8]{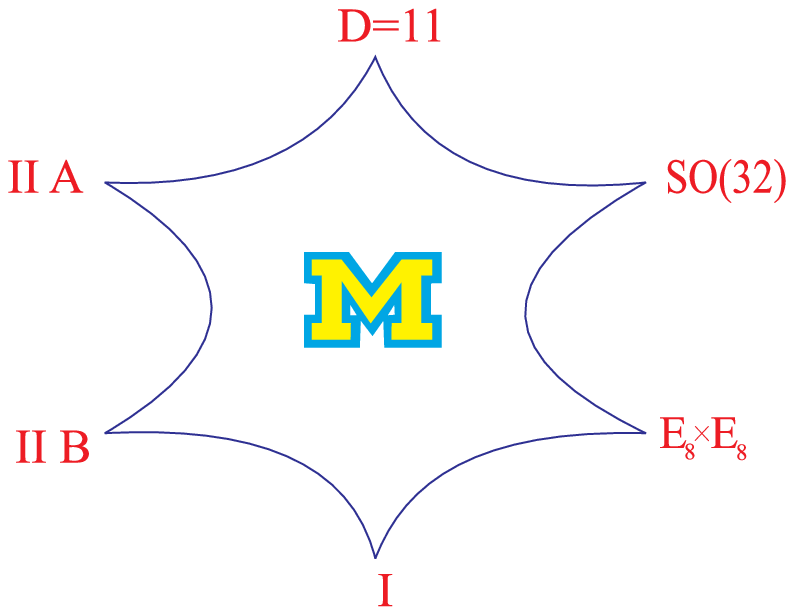}
\caption{M-theory subsumes eleven-dimensional supergravity and the five
ten-dimensional superstring theories.}\label{Mdiagram}\end{figure}

What may be less familiar to the general public is that
superstring theory has recently been superseded by a deeper and
more profound new theory, called ``M-theory'' \cite{Duff}.
M-theory involves membrane-like extended objects with two space
dimensions and five space dimensions that themselves live in a
universe with eleven spacetime dimensions \cite{Duff0} (ten space
and one time). As we shall see, it not only subsumes all of the
new ideas of superstring theory but also revives older ideas on
eleven-dimensional supergravity.  See Figure \ref{Mdiagram}.  New
evidence in favor of this theory is appearing daily on the internet
and represents the most exciting development\footnote{Indeed, several
of the stores in Ann Arbor changed their name to ``M-dens'' and many
students at the University of Michigan can be seen wearing M-theory
tee shirts, M-theory baseball caps and drinking from M-theory mugs.}
in the subject since 1984 when the superstring revolution first burst
on the scene.

\section{The fundamental constituents of matter}

Theoretical physicists like to ask the big questions: How did the
Universe begin?  What are its fundamental constituents?  What are the
laws of nature that govern these constituents?

\begin{figure}\centering\includegraphics[scale=0.4]{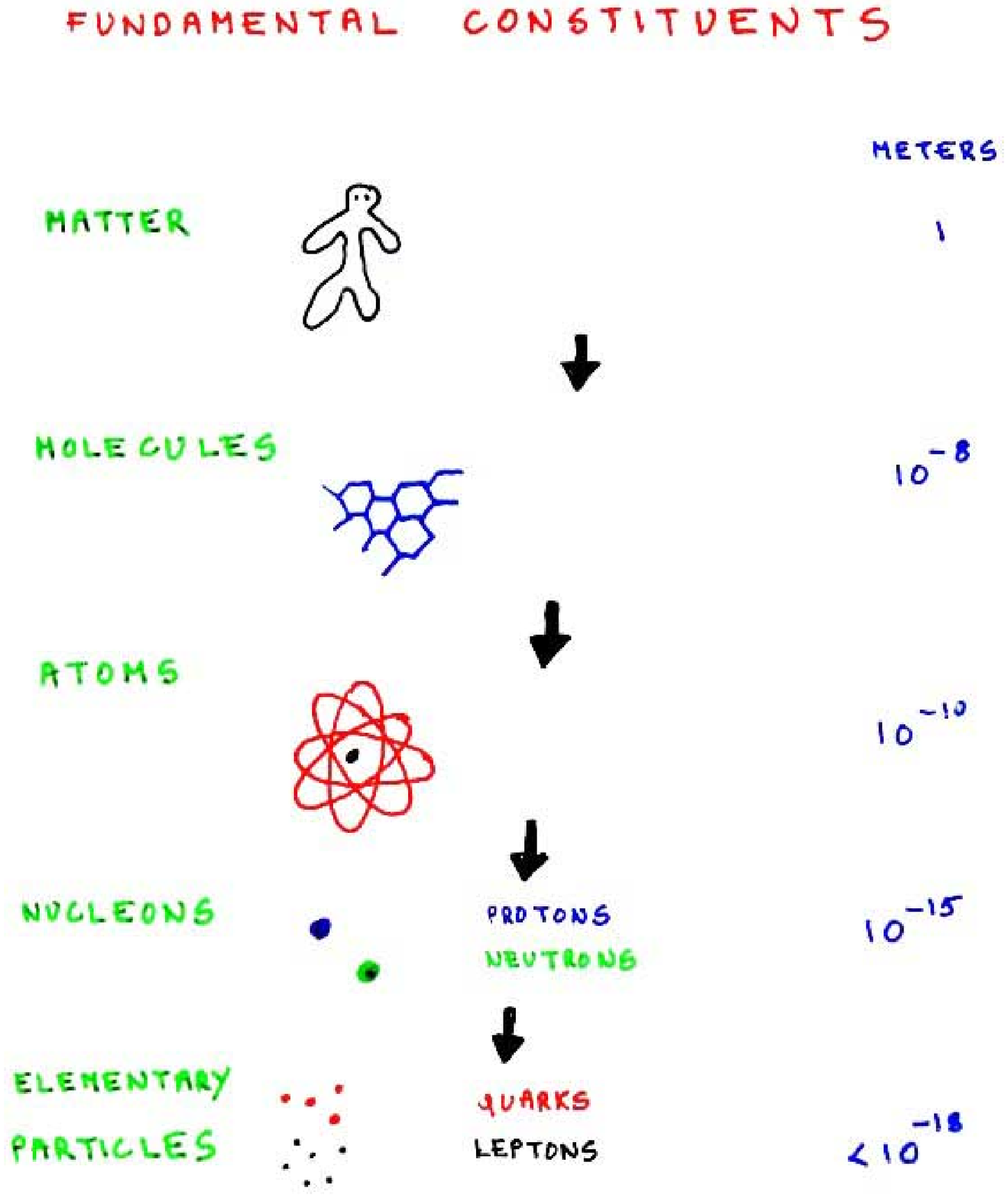}
\caption{Fundamental constituents.}\label{Constituents}\end{figure}

The smallest constituents of matter are, by definition, the {\it
elementary particles}.  But what is an elementary particle,
exactly? How do we know when we have reached the bottom line?
Well, it turns out to be easier to say what an elementary particle
is not. For example, as illustrated in Figure \ref{Constituents},
a human being (1 meter) is not elementary because he or she is
made out of molecules.  A molecule ($10^{-8}$ meters) is not
elementary because it is made out of atoms. An atom ($10^{-10}$
meters) is not elementary because it is made out of electrons
orbiting a nucleus which is made out of protons and neutrons
(collectively called {\it nucleons} ).  Nucleons ($10^{-15}$
meters) are not elementary because they are made out of smaller
constituents called {\it quarks}.  Finally, however, physicists
believe that these quarks, together with other particles called
{\it leptons}, of which the electron is an example, are indeed
truly elementary.  They appear to have no size of their own and,
as far as we can tell (which is down to about $10^{-18}$ meters),
behave like geometrical points.

Because of this, elementary particle physics is qualitatively
different from all other branches of science.  In his or her everyday
research, the biologist can take for granted all of chemistry; the
chemist can take for granted all of atomic physics; the atomic
physicist can take for granted all of nuclear physics and the nuclear
physicist can take for granted all of elementary particle physics.
Particle physicist are different, however, because they can take
nothing for granted.  There is no higher court of appeal.  Someone once
said that if you compare scientific
research to a game of chess, then most scientists are trying to become
masters of the game.  But the particle physicist is still trying to
figure out what the rules are!

Let us now take a preliminary look at the matter particles as displayed in
Table \ref{quarks}. Theoretical consistency, borne out by experiment,
requires that quarks and leptons come in families.  For reasons we do
not yet understand theoretically, the number of families is
exactly three.  Ordinary matter that we encounter on earth outside of
particle accelerators is composed only of the first family, which
consists of an electron neutrino with electric charge $Q=0$, an
electron with $Q=-1$, an {\it up} quark with $Q=+2/3$ and a {\it down}
quark with $Q=-1/3$.  These particles are denoted $\nu_{e}$, $e$, $u$
and $d$ respectively.  All members have a non-zero rest mass except
the neutrino which therefore travels at the speed of light.  Apart
from being heavier, the second and third families are identical
replicas of the first.  The second consists of a muon neutrino
$\nu_{\mu}$, a muon $\mu$, a {\it charm} quark $c$ and a {\it strange}
quark $s$.  The third consists of a tau neutrino $\nu_{\tau}$, a tau
$\tau$, a {\it top} quark $t$ and a {\it bottom} quark $b$.  The six
different types of quark are known as the six {\it flavors}.  However,
each quark also comes in three different {\it colors} labeled red,
green and blue after the three primary colors.  This is of course
just an analogy, but it is a good one because just we can form {\it
colorless} combinations from red, green and blue, so we can combine
colored quarks to form colorless bound states.  In fact, it seems to
be a fact of nature that the only particles we actually observe are
such colorless combinations.  In particular, we never observe free
quarks.  It is a prediction of quantum field theory that for every
particle there is an antiparticle with the same mass but opposite
charge.  So in addition to all the above particles, we also have their
antiparticles.  These are denoted by a bar, so that an anti-up quark,
for example would be $\overline{u}$.  Particles that were previously
believed to be elementary are now known to be built out of these basic
entities.  For example, the proton ($Q=1$) is a $uud$ combination, the
neutron ($Q=0$) is a $ddu$ and the positively charged pion ($Q=1$) is
a $u\overline{d}$.
\begin{table}
$$\matrix{I & {\nu}_e & e^- & u & d \cr
II & {\nu}_{\mu} & {\mu}^- & c & s \cr
III & {\nu}_{\tau} & {\tau}^- & t & b \cr
& Q=0 & Q=-1 & Q=2/3 & Q=-1/3 \cr}$$
\caption{Three families of quarks and leptons, with each quark coming in
three colors  (red, green and
blue).}
\label{quarks}
\end{table}

These fundamental building blocks of nature are held together by four
fundamental forces, listed in Table \ref{for} in order of diminishing
strength.
The strong nuclear force holds the protons and neutrons together inside the
atomic nucleus; the electromagnetic force is
responsible for all electric and magnetic phenomena and for light
(light is an electromagnetic wave): the weak nuclear
force is responsible for radioactivity, and lastly we have the
gravitational force which is most familiar in our
everyday lives.  As we see, the gravitational force is incredibly
weak compared to the other three forces and, for this reason, most
text-books will tell you that gravity is of no consequence at the
subatomic scale.  However, I will try to convince you that most text-books
are
wrong: not only will gravity prove to be important but all the
theoretical evidence suggests that it is actually at the root of
everything else.

\begin{figure}\centering\includegraphics[scale=0.8]{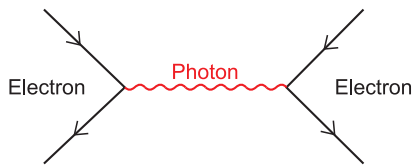}
\caption{Feynman diagram representing the force between two
electrons.}\label{Photon}\end{figure}

\begin{figure}\centering\includegraphics[scale=0.8]{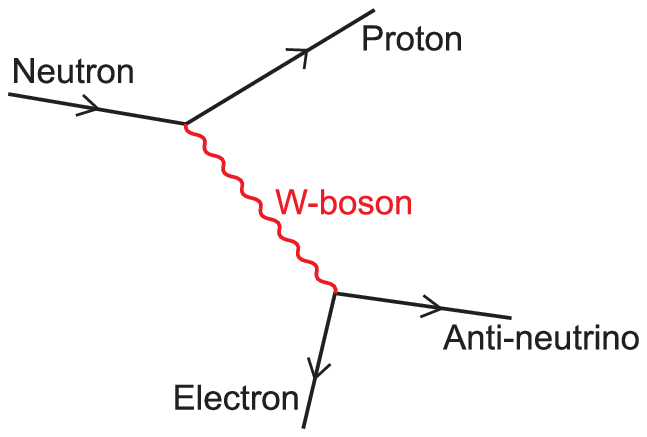}
\caption{Feynman diagram representing the radioactive decay of a
neutron.}\label{W}\end{figure}

 What exactly do we mean by a {\it force}?  Over the years, our answer to
this question has become more and more sophisticated.  Nowadays, we
employ quantum field theory, which is what we are led to when we try
and combine quantum mechanics with special relativity.  According to quantum
field theory, each force is associated with a {\it force carrier} .
For example, when an electron repels another electron, it does so by
exchanging the fundamental quantum of electromagnetism, the {\it
photon}, as in Figure \ref{Photon}.  By the way, this picture is an
example of what are called {\it Feynman diagrams}, named after Richard
Feynman, who invented them.  The study of the interactions between
electrons and photons is known as {\it quantum electrodynamics} or
{\it QED} and, in terms of the accuracy with which it agrees with
experiment, it is the most successful theory ever invented.  For
example, a measure of the electron's magnetism is the g-factor.
According to QED, it is given by
\begin{equation}
g/2=1.001159652190
\end{equation}
whereas the experimental value is
\begin{equation}
g/2=1.001159652193
\end{equation}
Because of this success, physicists have invented other force carriers to
account for the other forces.  The carriers of the strong force are known as
{\it gluons} and they act on quarks.  The study of the interactions between
colored quarks and gluons is known as {\it quantum chromodynamics} or {\it
QCD}.  The existence of the gluon was established indirectly in 1978.
The carriers of the weak force are known as W and Z bosons and,
together with the photon, they act on both quarks and leptons.  For
example, the phenomenon of radioactivity arise from the so-called
$\beta$-decay of a neutron into a proton, an electron and an
antineutrino as shown by the Feynman diagram in Figure \ref{W}.
This interaction is mediated by the exchange of a W boson.  The W
boson was discovered at CERN in 1982 and the Z boson, the following
year.  The collective study of the interactions of quarks and leptons
with photons, W and Z bosons is known as {\it electro-weak theory}.

Finally, physicists have coined the word {\it graviton} to describe
the force carrier of gravity, which acts on every other matter
particle and force carrier.  This is, at the moment, only a
hypothetical particle and is unlikely to be detected experimentally
for many years to come.

So far, we have not discussed the origin of mass within the Standard
Model.  How come the neutrino is massless but the electron is not?  How
come the photon is massless but the W is not?  It turns out that it is
necessary to introduce a third kind of particle, called the {\it Higgs
boson} after the Scottish theorist Peter Higgs, whose job it
is cleverly to give masses to the particles that need masses (like the
$e$ and the W) while not giving masses to those that do not (like the
$\nu$ and the $\gamma$).  Thus the Higgs is neither a force carrier
nor a matter particle.

Each elementary particle carries an intrinsic angular momentum or
{\it spin}, $s$, which can either be an integer $(s=0, 1, 2...)$, in
which case it is called a {\it boson}, or an odd half-integer
$(s=1/2, 3/2, 5/2...)$, in which case it is called a {\it fermion}.
The force carriers (gluons, photons, $W$ and $Z$) are all bosons with
$s=1$; the matter particles (quarks and leptons) are all fermions with
$s=1/2$; the Higgs is an $s=0$ boson. Bosons and fermions behave very
differently.  For example, fermions obey the {\it exclusion principle} of
Wolfgang Pauli, which states that no two fermions can occupy the same
quantum state, whereas bosons do not.  They are said to obey {\it
opposite statistics}.

The {\it Standard Model} is the name we give to the collection of
strong, weak and electromagnetic interactions.  It describes the
interaction of the matter particles, force carriers and Higgs.  It is
consistent with every experiment that has yet been
performed\footnote{Recent experiments
have indicated that the neutrino has a very small but non-zero mass,
but this would require    only relatively mild modifications to the Standard
Model.}, but it also predicts other effects which have not yet been
seen.  In particular, there is still no direct experimental evidence
for the Higgs boson.  An important goal of future experiments at Fermilab
in Chicago and the Large Hadron Collider at CERN in Geneva, will be to
hunt for the Higgs.

\begin{figure}\centering\includegraphics[scale=0.7]{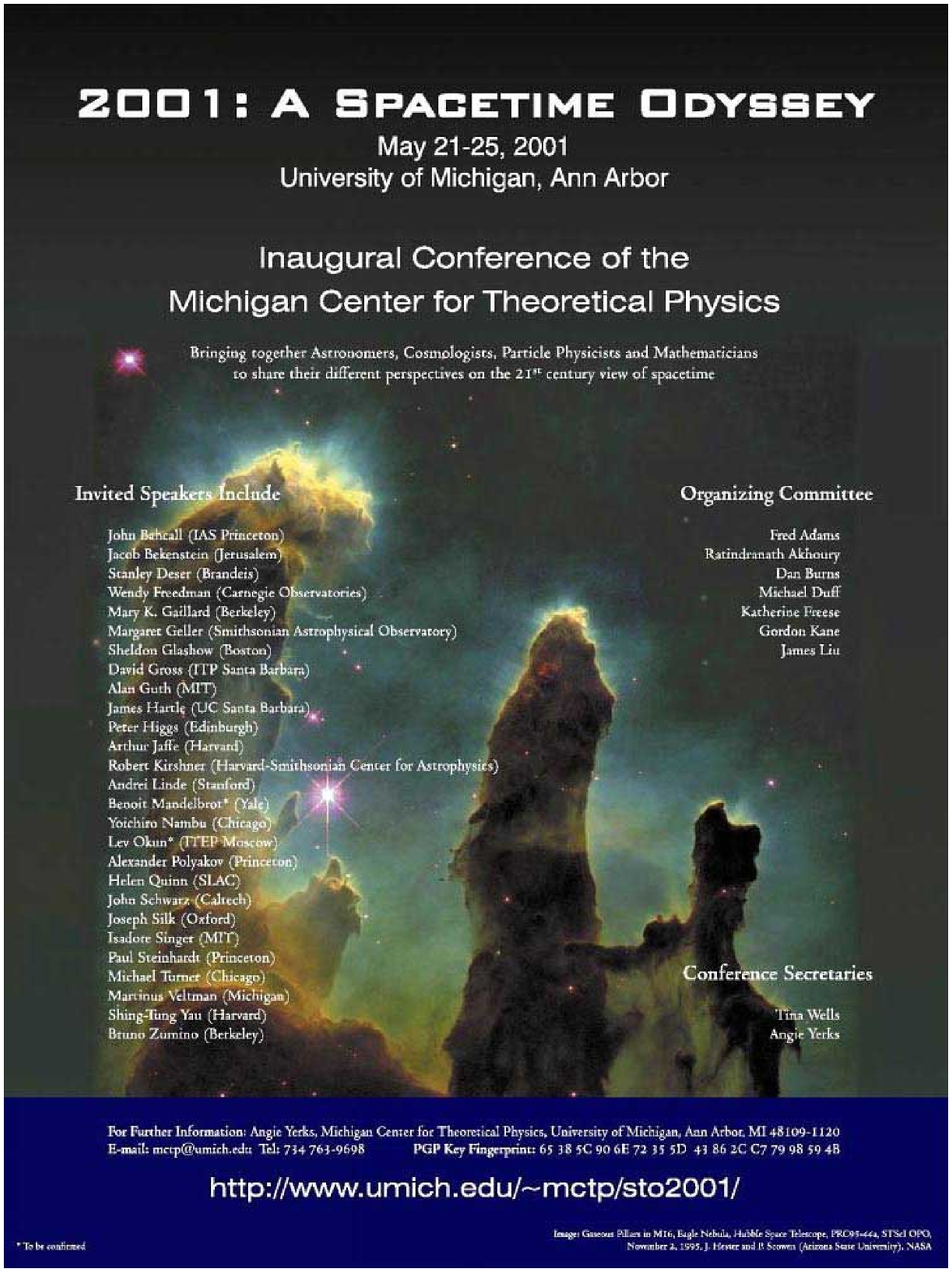}
\caption{2001: A Spacetime Odyssey.}\label{Odyssey}\end{figure}

Three of the heroes of the Standard Model, Sheldon Glashow (who
won the 1979 Nobel Prize for explaining how to unify the weak and
electromagnetic forces), Martinus Veltman (who won the 1999 Nobel
prize for showing how the theory is physically and mathematically
consistent) and Peter Higgs (without whose boson everything in the
universe would be massless, and who therefore bears a heavy
responsibility, in more ways than one) will be speaking at ``2001:
A Spacetime Odyssey'', the Inaugural Conference of the Michigan
Center for Theoretical Physics to be held here in Ann Arbor in May
\cite{Odyssey}. See Figure \ref{Odyssey}.

\begin{table}[h]
\centering $\begin{array}{lcll}
FORCE & STRENGTH & CARRIER & ACTS~ON \\
~&~&~&\\
strong & 1 & gluon & quarks \\
electromagnetic & 10^{-2} & photon & quarks~and ~leptons\\
weak & 10^{-5} &W~and~Z~bosons & quarks~and~leptons\\
gravity & 10^{-38} & graviton &everything
\end{array}$
\caption{The fundamental forces.}
\label{for}
\end{table}

\section{What about gravity?}

\begin{figure}\centering\includegraphics[scale=0.8]{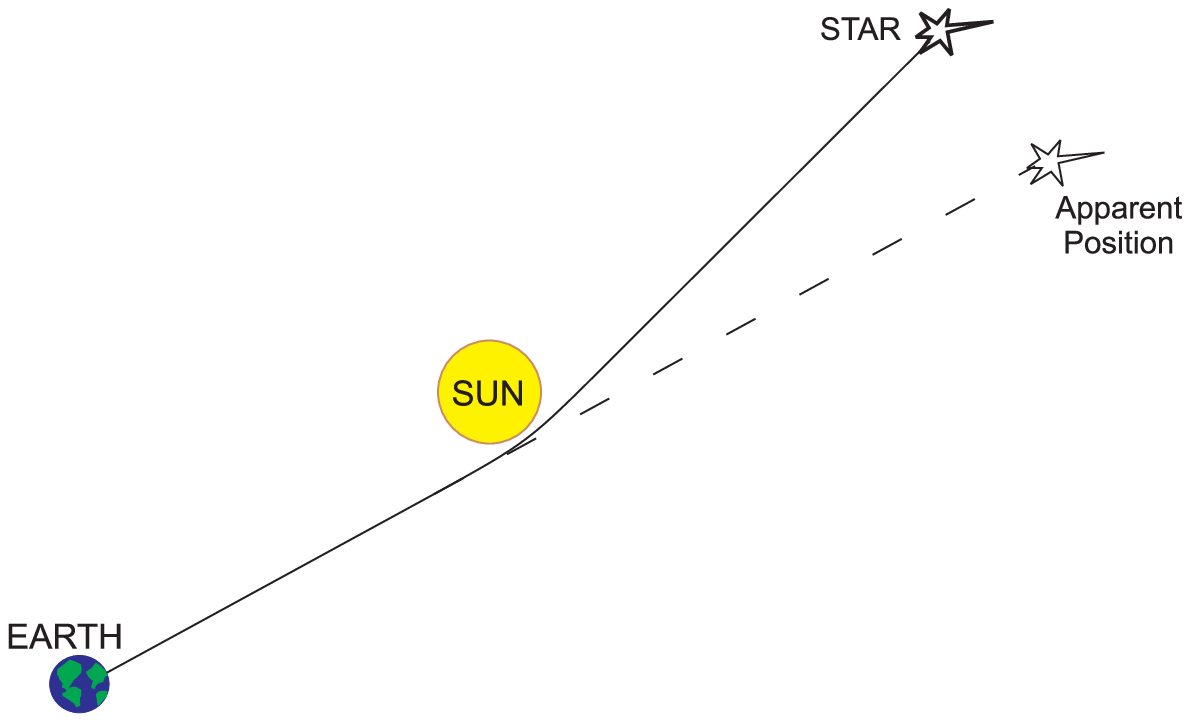}
\caption{The bending of light by the Sun.}\label{Sun}\end{figure}

Why is it so difficult to incorporate the fourth force of gravity into
quantum
physics?  Our modern understanding of the gravitational force dates back to
Einstein's 1916 theory of general relativity which asserts that the laws of
physics should be the same to {\it all} observers, not merely those in
uniform
relative motion as in special relativity.  According to
this picture, gravity is not a {\it force}  at all. All bodies are following
straight line trajectories, but in a {\it curved spacetime} .  (Note that it
is
not merely three-dimensional space which is curved, but the four-dimensional
spacetime continuum.)  A classic illustration of this idea is provided by
the
bending of light by the Sun, as shown in Figure \ref{Sun}.  At first
sight, these classical geometrical ideas seem to be at odds with the
quantum picture of force-carrying gravitons.  In recent times, however,
we have come to appreciate that these two pictures are complementary.
Indeed, the emphasis now is to explain the other three forces in
geometrical terms as well.

The goal of modern theoretical physics is to find an
all-embracing ``Theory of Everything'' that would unite all four forces.
But if current thinking about M-theory is correct, this will require three
radical ideas:

1) Supersymmetry

2) Extra spacetime dimensions

3) Extended objects

\section{Supersymmetry}

Central to the understanding of modern theories of the fundamental forces
is the idea of {\it symmetry}: under certain changes in the way we
describe the basic quantities, the laws of physics are nevertheless seen
to remain unchanged. For example, the result of an experiment should be
the same whether we perform it today or tomorrow; this symmetry is called
{\it time translation invariance}. It should also be the same before and
after rotating our experimental apparatus; this symmetry is called {\it
rotational invariance}. Both of these are examples of {\it spacetime
symmetries}. Indeed, Einstein's general theory of relativity is based on
the requirement that the laws of physics should be invariant under {\it
any} change in the way we describe the positions of events in spacetime.
In the Standard Model of the strong, weak and electromagnetic forces there
are other kinds of {\it internal} symmetries that allow us to change the
roles played by different elementary particles such as electrons and
neutrinos, for example.  In Grand Unified Theories, which have not yet
received the same empirical support as the Standard Model, the laws
remain unchanged even when we exchange the roles of the {\it quarks}
and electrons.  Thus it is that the greater the unification, the
greater the symmetry required.  The Standard Model symmetry replaces
the three fundamental forces: strong, weak and electromagnetic, with
just two: the strong and electroweak.  Grand unified symmetries
replace these two with just one strong-electroweak force.  In fact, it
is not much of an exaggeration to say that the search for the ultimate
unified theory is really a search for the right symmetry.

At this stage, however, one might protest that some of these internal
symmetries fly in the face of experience. After all, the electron is very
different from a neutrino: the electron has a non-zero mass whereas the
neutrino is massless.  Similarly, the electrons which orbit the
atomic nucleus are very different from the quarks out of which the
protons and neutrons of the nucleus are built.  Quarks feel the strong
nuclear force which holds the nucleus together, whereas electrons do
not.  These feelings are, in a certain sense, justified: the world we
live in does not exhibit the symmetries of the Standard Model nor
those of Grand Unified Theories.  They are what physicists call
``broken symmetries''.  The idea is that these theories may exist in
several different {\it phases}, just as water can exist in solid,
liquid and gaseous phases.  In some of these phases the symmetries are
broken but in other phases, they are exact.  The world we inhabit
today happens to correspond to the broken-symmetric phase, but in
conditions of extremely high energies or extremely high temperatures,
these symmetries may be restored to their pristine form.  The early
stages of our universe, shortly after the Big Bang, provide just such
an environment.  Looking back further into the history of the
universe, therefore, is also a search for greater and greater
symmetry.  The ultimate symmetry we are looking for may well be the
symmetry with which the Universe began.

$M$-theory, like string theory before it, relies crucially on the idea,
first put forward in the early 1970s, of a spacetime {\it supersymmetry}
which exchanges bosons and fermions.  Unbroken
supersymmetry would require that every elementary particle we know of
would have an unknown super-partner with the same mass but obeying the
opposite statistics: for each boson there is a fermion; for each fermion a
boson.  Spin $1/2$ quarks partner spin $0$ {\it squarks}, spin $1$ photons
partner spin $1/2$ {\it photinos}, and so on. In the world we inhabit, of
course, there are no such equal mass partners and bosons and fermions seem
very different. Supersymmetry, if it exists at all, is clearly a broken
symmetry and the new supersymmetric particles are so heavy that they have
so far escaped detection. At sufficiently high energies, however,
supersymmetry may be restored.  Another challenge currently facing
high-energy experimentalists at Fermilab and CERN is the search for
these new supersymmetric particles.  The discovery of supersymmetry
would be one of the greatest experimental achievements and would
completely revolutionize the way we view the physical
world\footnote{This will present an interesting dilemma for those
pundits who are predicting the {\it End of Science} on the grounds
that all the important discoveries have already been made.
Presumably, they will say ``I told you so'' if supersymmetry is not
discovered, and ``See, there's one thing less left to discover'' if it
is.}.

Symmetries are said to be {\it global} if the changes are the same
throughout spacetime, and {\it local} if they differ from one point to
another. The consequences of {\it local} supersymmetry are even more
far-reaching: it predicts gravity. Thus if Einstein had not already
discovered General Relativity, local supersymmetry would have forced
us to invent it. In fact, we are forced to a {\it supergravity} in which the
graviton, a spin $2$ boson
that mediates the gravitational interactions, is partnered with a spin
$3/2$ {\it gravitino}. This is a theorist's dream because it confronts the
problem from which both general relativity and grand unified theories shy
away: neither takes the other's symmetries into account. Consequently,
neither is able to achieve the ultimate unification and roll all four
forces into one. But local supersymmetry offers just such a possibility,
and it is this feature above all others which has fuelled the theorist's
belief in supersymmetry in spite of thirty years without experimental
support.

Supergravity has an even more bizarre feature, however, it places an upper
limit of {\it eleven} on the dimension of spacetime! There are
two different ways to see this. The original explanation relies on the
belief that there are no consistent ways to describe massless particles with
spins greater than two, the spin of the graviton.  Yet the mathematics
of supersymmetry says that spins greater than two (and more than one
graviton) would have to appear in dimensions greater than eleven, thus
leading to a contradiction.  The second explanation, to which we shall
return in
section \ref{Supermembranes}, does not rely on the absence of higher spins,
but
says the maximum dimension in which there exists supersymmetric
objects such as particles, strings and membranes is eleven, where in
fact we find a supermembrane.

We are used to the
idea that space has three dimensions: height, length and breadth; with time
providing the fourth dimension of {\it spacetime}.  Indeed this is the
picture that Einstein had in mind in 1916 when he proposed general
relativity.  But in the early 1920's, in their attempts to unify
Einstein's gravity and Maxwell's electromagnetism, Theodor Kaluza and
Oskar Klein suggested that spacetime may have a hidden fifth dimension.
This Kaluza-Klein theory, and its higher dimensional generalizations,
are thus tailor-made for supersymmetry.

\section{The fifth dimension}

In 1864 James Clerk Maxwell introduced the equations that describe the
electromagnetic field.  Albert Einstein later realized in 1905 that
Maxwell's equations obey the principle of special relativity, that the
laws of physics should be the same to all observers who are in uniform
relative motion.  In special relativity, which treats time as a fourth
dimension, time $t$ and the space coordinates $x,y,z$ are collectively
denoted $x^{\mu}$ where the index $\mu$ runs over $0,1,2,3$.  The $0$
refers to time and $1,2,3$ to the three space coordinates $(x,y,z)$:
\begin{equation}
(x^{0},x^{1},x^{2},x^{3})=(t,x,y,z)
\end{equation}
In this modern notation, Maxwell's electromagnetic field also has four
components collectively denoted $A_{\mu}(x)$.  The field depends on its
position in spacetime and so is a function of the
$4$ spacetime coordinates $x^{\mu}$.

In 1916, Einstein introduced the principle of general relativity, that
the laws of physics should be the same to {\it all} observers. This
necessitates a gravitational field with two indices,
$g_{\mu\nu}(x)$, which also has the geometrical interpretation of a metric
tensor, the quantity that
describes the infinitesimal distance $ds$ between two points in
four-dimensional spacetime.
\begin{equation}
ds^{2}=g_{\mu\nu}(x)dx^{\mu}dx^{\nu}
\end{equation}
Note that the Euclidean
geometry of flat spacetime must be replaced by the Riemannian geometry of
curved spacetime for which the metric tensor is itself a function of the
spacetime coordinates.

In 1919, therefore, in the search for a unified theory, it was natural
to attempt to combine Maxwell's electromagnetism with Einstein's
gravity, since the other two nuclear forces were not as well understood.
This the German-Polish mathematician Theodor Kaluza was able to do by the
ingenious device of postulating a fifth dimension with coordinate
$\theta$.  The five coordinates are denoted collectively $x^{M}$ where
the index $M$ runs over $0,1,2,3,4$.
\begin{equation}
(x^{0},x^{1},x^{2},x^{3},x^{4})=(t,x,y,z,\theta)
\end{equation}

He imagined a five
dimensional Riemannian geometry with metric tensor ${\hat g}_{MN}(x)$ which
describes the infinitesimal distance $d{\hat s}$ between two points in
this five-dimensional spacetime
\begin{equation}
d{\hat s}^{2}={\hat g}_{MN}(x)dx^{M}dx^{N}
\end{equation}
He then made a $4+1$ split

\begin{eqnarray}
{\hat g}_{MN}&=& \left(
\begin{array}{cc}
g_{\mu\nu}+{\Phi} A_{\mu}A_{\nu}&{\Phi}A_{\mu}\\
{\Phi}A_{\nu}&{\Phi}
\end{array}
\right)\nonumber\\
\label{kaluza}
\end{eqnarray}
and identified $g_{\mu\nu}(x)$ with Einstein's gravitational field and
$A_{\mu}(x)$ with Maxwell's electromagnetic field.  This was all before
the advent of quantum field theory, but nowadays we would refer to
$g_{\mu\nu}$ as the spin $2$ graviton, $A_{\mu}$ as the spin $1$
photon and $\Phi$ as the spin $0$ dilaton\footnote{This field was
considered an embarrassment in 1919, and was (inconsistently) set equal
to zero.  However, it was later revived and subsequently stimulated
Brans-Dicke theories of gravity.  The dilaton also plays a crucial
role in superstring and M-theory.}.

Of course, it is not enough to call $A_{\mu}$ by the name photon, we
must demonstrate that it obeys the right equations and here we see the
Kaluza miracle at work.  If we substitute (\ref{kaluza}) into the five
dimensional field equations, not only do we recover the correct
Einstein equations for $g_{\mu\nu}$, but also the Maxwell equations
for $A_{\mu}$.  So Maxwell's theory of electromagnetism was seen to be
a consequence of Einstein's general relativity, provided you are
willing to buy the idea of an extra spacetime dimension.

By the way, Kaluza's son, who still teaches mathematics in Germany,
recalls  that his father belonged to that school of theoreticians who
believed that everything in nature could be derived from pure thought
without
the need for experiment. Consequently, he learned to swim from
a textbook. He would lie on the living room couch with the book in one hand
and practice his strokes with the other.  He then proceeded to walk to the
nearest lake and swim across it. Personally, I find it easier to
believe in a fifth dimension than in the veracity of that particular story.

\begin{figure}\centering\includegraphics[scale=0.8]{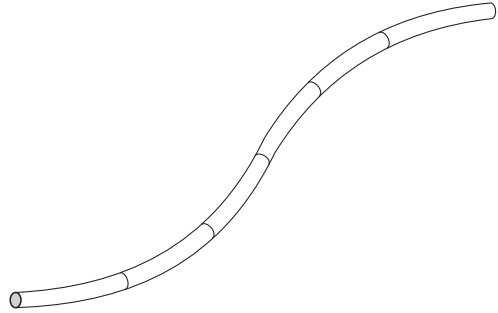}
\caption{A garden hose. At large distances, it looks like a line; closer
inspection reveals that at each point of the line there is a
little circle.}\label{Hose}\end{figure}

Attractive though Kaluza's idea was, it suffered from two obvious drawbacks.
First, although the indices were allowed to range over $0,1,2,3,4$, for no
very
good reason the dependence of the fields on the extra
coordinate $\theta$ was suppressed. Secondly, if there is a fifth dimension
why
haven't we seen it? The resolution of both these problems was supplied by
Oskar Klein\footnote{Not to be confused (as the {\it Oxford Dictionary
of Physics} is) with Felix Klein (no relation), inventor of the Klein
bottle.} in 1926 \cite{Klein1,Klein2}.  Klein insisted on treating the
extra dimension seriously but assumed the fifth dimension to have
circular topology so that the coordinate $\theta $ is periodic, $0\leq
\theta \leq 2\pi$.  It is difficult to envisage a spacetime with this
topology but a simpler two-dimensional analogy is provided by a garden
hose: at large distances it looks like a line but closer inspection
reveals that at every point on the line there is a little circle.  So
it was that Klein suggested that there is a little circle at each
point in four-dimensional spacetime.  See Figure \ref{Hose}.

The periodicity in $\theta$ means that the fields ${\hat g}_{MN}(x,\theta)$
may
be expanded in the form
\begin{equation}
 {\hat g}_{MN}(x,\theta)= \sum_{n=-\infty}^{n=\infty}
 {\hat g}_{MN(n)}(x)exp({in\theta}),
 \label{fourier}
\end{equation}
The $n=0$ modes in (\ref{fourier}) are just Kaluza's graviton, photon and
dilaton.  If we now include the $\theta$-dependence via the $n\neq0$ modes,
however, we find an infinite tower of charged, massive spin $2$
particles with charges $e_n$ given by
\begin{equation}
e_n=ne,
\label{charge}
\end{equation}
and masses $m_n$ given by
\begin{equation}
m_n=|n|/R
\label{mass}
\end{equation}
where $R$ is the radius of the circle and
\begin{equation}
e^{2}=16\pi G/R^{2} \label{charge1}
\end{equation}
where $G$ is Newton's constant of gravitation.
Thus Klein explained (for the first time) the empirical fact that
all observed particles come with an electric charge which is an integer
multiple of a fundamental charge $e$, in other words, why electric
charge is {\it quantized}.  Of course, if we identify this fundamental
unit of charge with the charge on the electron, then we are forced to
take the radius of the circle to be very small: the Planck size
$10^{-35}$ meters; much smaller than the $10^{-18}$ meters achievable
by current particle accelerators.  This satisfactorily accords with
our everyday experience of living in four spacetime dimensions.

\section{Oskar Klein}

\begin{figure}\centering\includegraphics*[scale=0.5]{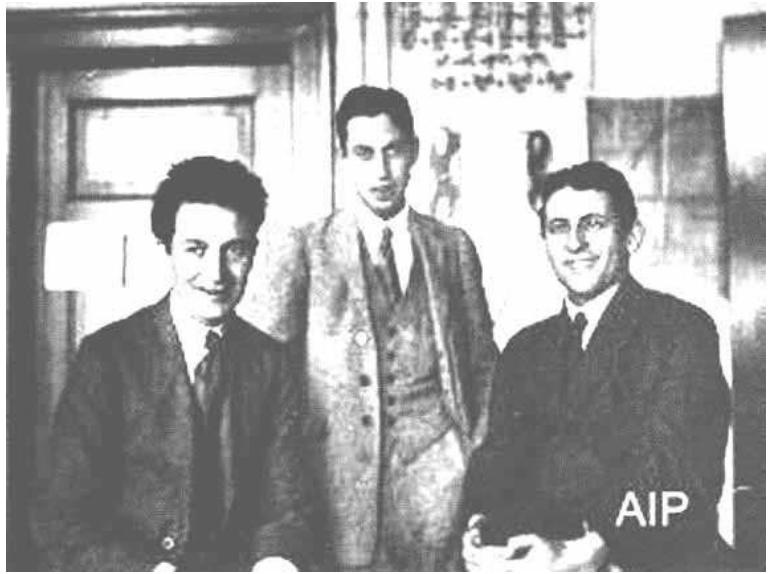}
\caption{Left to right: Oskar Klein, George Uhlenbeck and Samuel
Goudsmit (American Institute of
Physics).}\label{Uhlenbeck-Goudsmit}\end{figure}

As the quote on the cover page shows, the idea for a fifth dimension
came to Klein during his stay as an assistant professor at the
University of Michigan, 1923-25.  He was hired by H.  M.  Randall on the
recommendation of
the father of quantum theory, Niels Bohr, with whom he had been
working in Copenhagen.  He was a contemporary of two other famous
Michigan faculty members, George Uhlenbeck and Samuel Goudsmit,
discoverers of electron spin\footnote{Visitors to the Michigan Center
for Theoretical Physics (http://www.umich.edu/\~{}mctp/) are invited to view
a collection of memorabilia and  photographs of Klein (with Bohr, Goudsmit,
Uhlenbeck and others) in the Oskar Klein Conference Room, 3481
Randall Laboratory.}.  See Figure \ref{Uhlenbeck-Goudsmit}.

\begin{figure}\centering\includegraphics[scale=1]{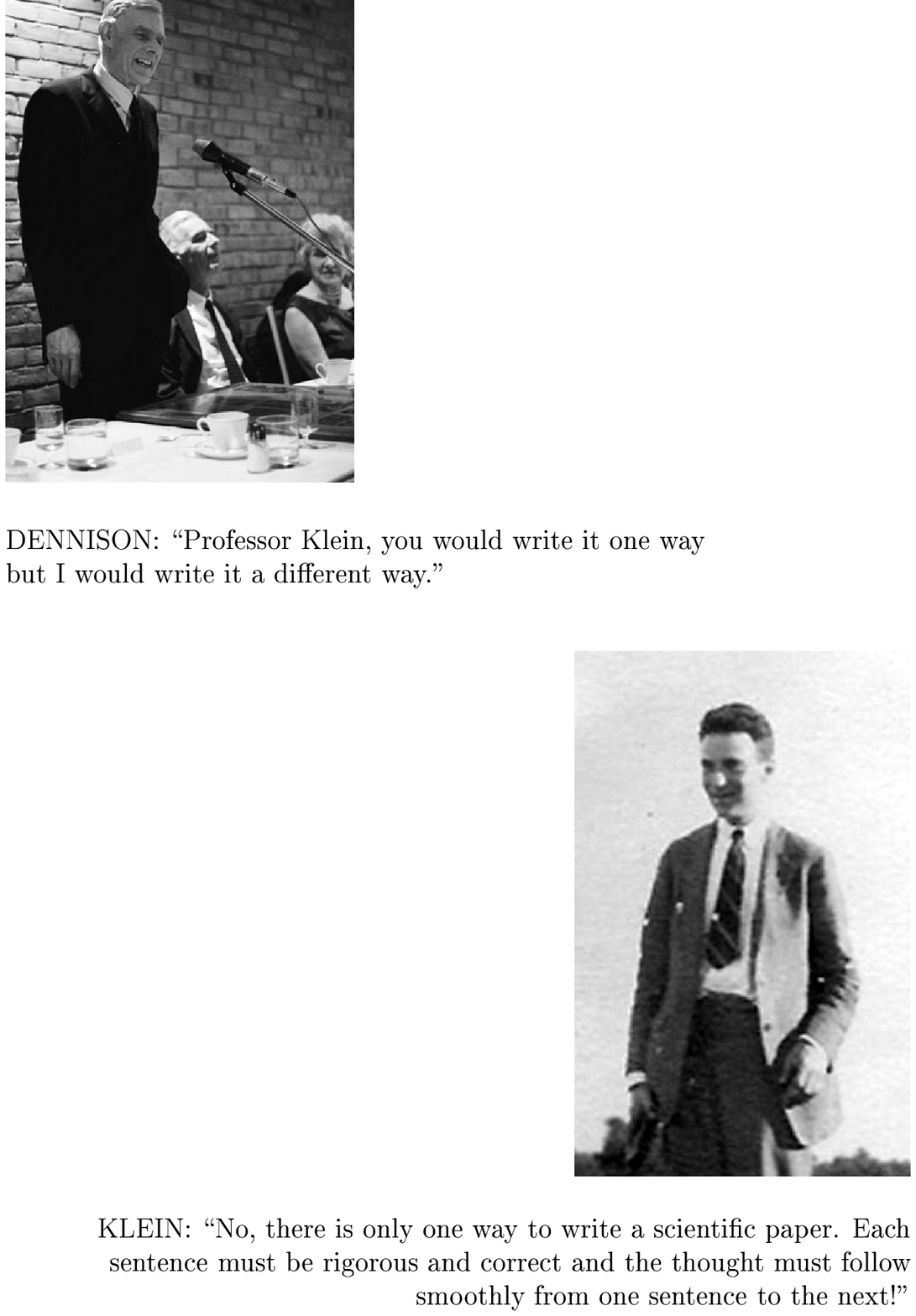}
\caption{Dennison's recollection of his PhD
thesis.}\label{Dennison}\end{figure}

Klein was also the thesis advisor of another distinguished Michigan
faculty member, David Dennison, discoverer of proton spin.  As
Dennison recalls \cite{Dennison}, Klein proved to be hard task master.
See Figure \ref{Dennison}.

Klein was not at first aware of Kaluza's earlier idea
but learned about it from Wolfgang Pauli to whom he had shown his
manuscript. Ever the gentleman, he generously acknowledges Kaluza's
prior claim, but as we have noted above, it was Klein who really
took seriously the extra dimension with all its implications, and by
assigning it the topology of a circle, provided the first explanation
of electric charge quantization.

\begin{figure}\centering\includegraphics[scale=0.35]{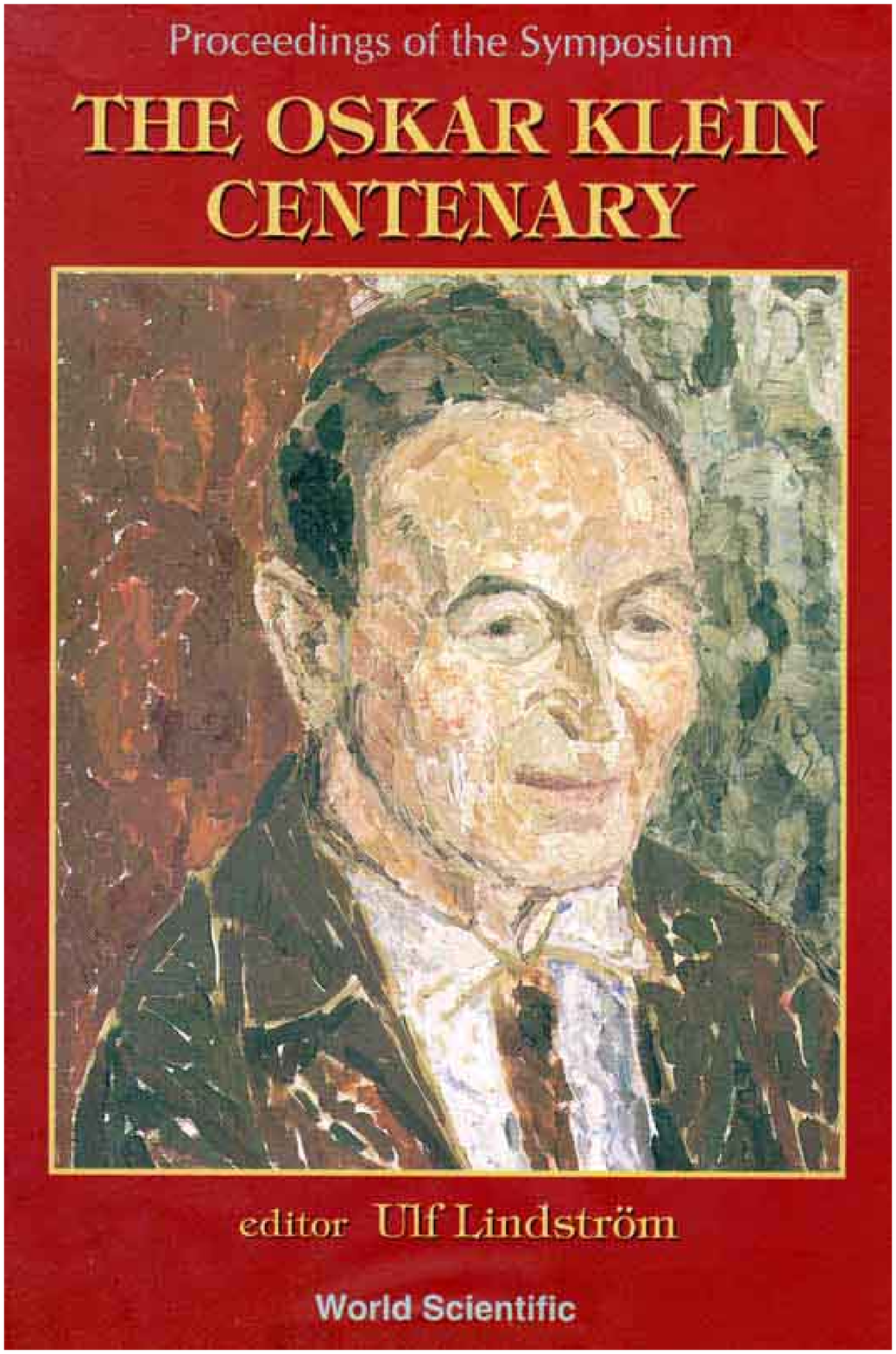}
\caption{The Oskar Klein Centenary, 1994}\label{Centenary}\end{figure}

Klein was born in Sweden in 1894 and in 1994 the Nobel Committee of
the Royal Swedish Academy of Sciences organized The Oskar Klein
Centenary Symposium in Stockholm, at which I was privileged to
deliver a review of Kaluza-Klein theory \cite{Duff1}.  See Figure
\ref{Centenary}.  So I feel doubly privileged to be delivering this Oskar
Klein
Professorship Inaugural Lecture here today.

Klein was responsible for many other important discoveries,
including the Klein paradox, the Klein-Gordon equation and the
Klein-Nishina formula.  A discussion of these and more on his life and
times may be found in the recollections of that able historian of
physics Abraham Pais \cite{Pais1,Pais2} and of my friend and colleague
Stanley Deser \cite{Deser} who is married to Klein's daughter Elsbeth.

\section{Eleven dimensional supergravity}

The Kaluza-Klein idea was forgotten for many years but was revived in the
early 1980s when it was realized by Eugene Cremmer, Bernard Julia and
Joel Scherk from the Ecole Normale in Paris that supergravity not only
permits up to seven extra dimensions, but in fact takes its simplest and
most elegant form when written in its full eleven-dimensional glory.
Moreover, the kind of four-dimensional picture we end up with depends on
how we curl up or {\it compactify} these extra dimensions: maybe seven of
them would
allow us to derive, a la Kaluza-Klein, the strong and weak forces as well
as the electromagnetic \cite{Freedman}.  The four dimensional theory we end
up after such a compactification describes a spin 2 graviton and spin
3/2 gravitinos interacting with a collection of spin 1, spin 1/2 and
spin 0 particles.  This collection will depend on the choice of the
compact seven-dimensional space.  So the question was whether there
was a choice that yielded the gluons, W and Z-bosons, photons, quarks,
leptons and Higgs of the Standard Model.  In the end, however, eleven
dimensional supergravity fell out of favor for a couple of reasons.

First, an important feature of the real world which is incorporated into
both the Standard Model and Grand Unified Theories is that Nature is
{\it chiral}: the weak nuclear force distinguishes between right and
left.  However, as emphasized by Edward Witten of the Institute for
Advanced Study in Princeton, among others, it is impossible
via conventional Kaluza-Klein techniques to generate a chiral theory
from a non-chiral one and unfortunately, eleven-dimensional
supergravity, in common with any {\it odd}-dimensional theory, is
itself non-chiral.

Secondly, despite its extra dimensions and despite its supersymmetry,
eleven-dimensional supergravity is still a {\it quantum field theory}
and runs into the problem from which all such theories suffer: the
quantum mechanical probability for certain processes yields the answer
{\it infinity}, signalling a breakdown of the theory. There is a natural
energy scale associated with any quantum theory of gravity.  Such a
theory combines three ingredients each with their own fundamental
constants: Planck's constant $h$ (quantum mechanics), the velocity of
light $c$ (special relativity) and Newton's gravitational constant $G$
(gravity).  From these we can form the so-called Planck mass
$m_P=\sqrt{hc/G}$, equal to about $10^{-8}$ kilograms, and the Planck
energy $m_Pc^2$, equal to about $10^{19}$ GeV.  (GeV is short for
giga-electron-volts=$10^9$ electron-volts, and an electron-volt is the
energy required to accelerate an electron through a potential
difference of one volt.) From this we conclude that the energy at
which Einstein's theory, and hence eleven-dimensional supergravity,
breaks down is the Planck energy.  On the scale of elementary particle
physics, this energy is enormous\footnote{For this reason,
incidentally, the {\it End of Science} brigade like to claim that,
even if we find the right theory of quantum gravity, we will never be
able to test it experimentally!  As I will argue shortly, however,
this view is erroneous.} : the world's most powerful particle
accelerators can currently reach energies of only $10^{4}$ GeV.  So it
seemed in the early 1980s that we were looking for a fundamental
theory which reduces to Einstein's gravity at low energies, which
describes Planck mass particles and which is supersymmetric.  Whatever
it is, it cannot be a quantum field theory because we already know all
the supersymmetric ones and they do not fit the bill.

\section{Ten-dimensional superstrings}

For both these reasons, attention turned to ten-dimensional
superstring theory. The idea that the fundamental stuff of the
universe might not be pointlike elementary particles, but rather
one-dimensional strings had been around from the early 1970s. Just
like violin strings, these relativistic strings can vibrate and
each elementary particle: graviton, gluon, quark and so on, is
identified with a different mode of vibration. However, this means
that there are {\it infinitely many} elementary particles.
Fortunately, this does not contradict experiment because most of
them, corresponding to the higher modes of vibration, will have
masses of the order of the Planck mass and above and will be
unobservable in the direct sense that we observe the lighter ones.
Indeed, an infinite tower of Planck mass states is just what the
doctor ordered for curing the non-renormalizability disease. In
fact, because strings are {\it extended}, rather than pointlike,
objects, the quantum mechanical probabilities involved in string
processes are actually {\it finite}. Moreover,  when we take the
{\it low-energy limit} by eliminating these massive particles
through their equations of motion, we recover a ten-dimensional
version of supergravity which incorporates Einstein's gravity. Now
ten-dimensional theories, as opposed to eleven-dimensional ones,
also admit the possibility of {\it chirality}. The reason that
everyone had still not abandoned eleven-dimensional supergravity
in favor of string theory, however, was that the realistic-looking
Type $I$ string, which seemed capable of incorporating the
Standard Model of particle physics, seemed to suffer from
inconsistencies or {\it anomalies}, whereas the consistent
non-chiral Type $IIA$ and chiral Type $IIB$ strings did not seem
realistic.

\begin{figure}\centering\includegraphics[scale=0.4]{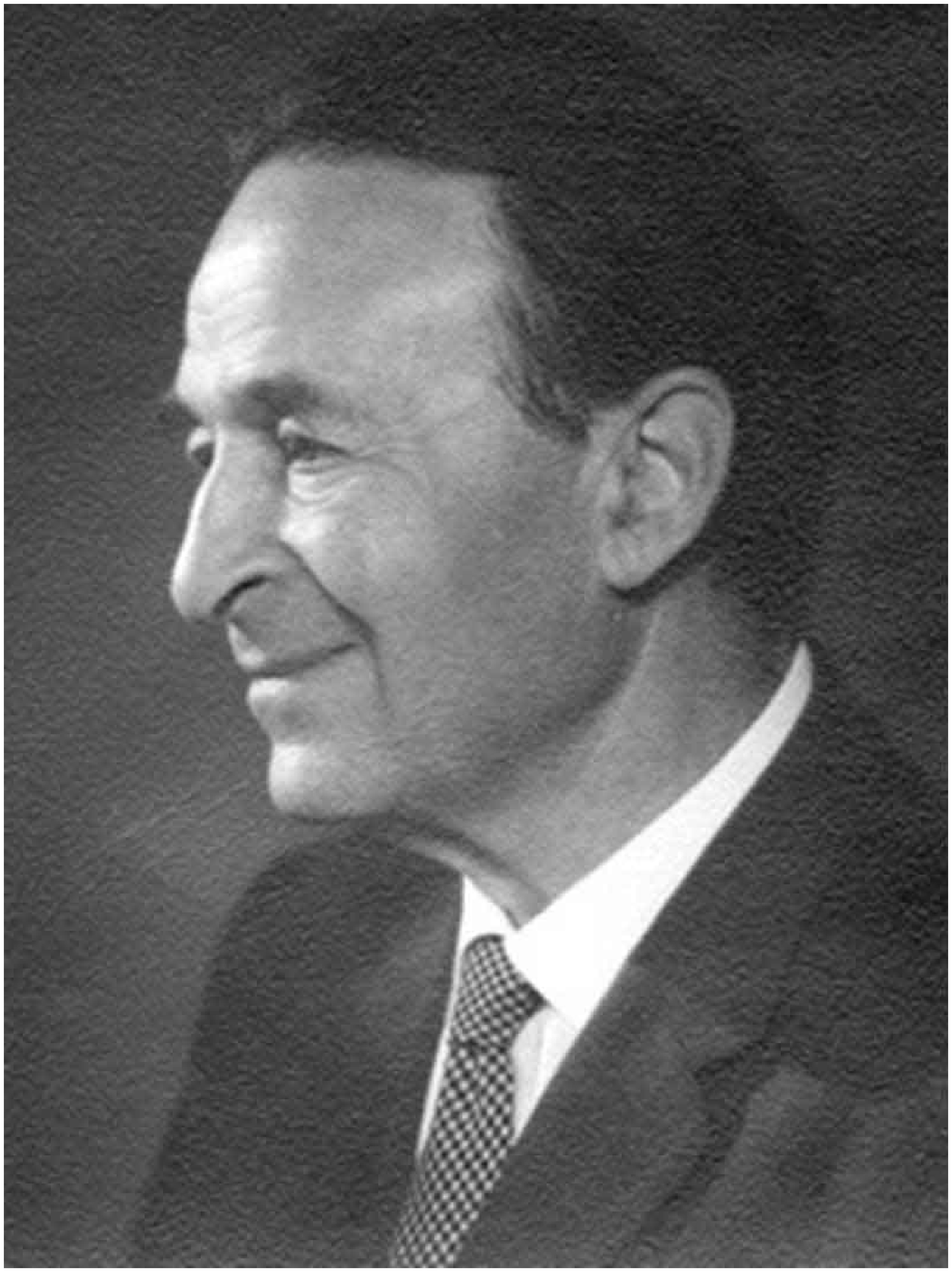}

\begin{flushleft} {\it A study of the history of science-not the
history of philosophy-shows that the natural attitude of a
scientist is to be inspired by the great predecessors, just as
they themselves were by their predecessors, but always taking the
liberty of doubt when there are reasons for doubt} \end{flushleft}

\begin{flushright}
\medskip Oskar Klein, ``From My Life
in Physics'' \end{flushright}
 \caption{The liberty of
doubt}\label{Klein}\end{figure}

Then came the September 1984 superstring revolution. First, Michael Green
from Queen Mary College, London, and John Schwarz from the California
Institute of Technology showed that the Type $I$ string was free of
anomalies provide the group was uniquely $SO(32)$ where O(n) stands
for {\it orthogonal} $n \times n$ matrices.  They suggested that a
string theory based on the exceptional group $E_8 \times E_8$ would
also have this property.  Next, David Gross, Jeffrey Harvey, Emil
Martinec and Ryan Rohm from Princeton University discovered a new kind
of heterotic (hybrid) string theory based on just these two groups:
the $E_8 \times E_8$ heterotic string and the $SO(32)$ heterotic
string, thus bringing to {\it five} the number of consistent string
theories.  Thirdly, Philip Candelas from the University of Texas,
Austin, Gary Horowitz and Andrew Strominger from the University of
California, Santa Barbara and Witten showed that these heterotic
string theories admitted a Kaluza-Klein compactification from ten
dimensions down to four.  The six-dimensional compact spaces belonged
to a class of spaces known to the pure mathematicians as {\it
Calabi-Yau manifolds}.  The resulting four-dimensional theories
resembled quasi-realistic Grand Unified Theories with chiral
representations for the quarks and leptons!  So at last we had a
consistent quantum theory of gravity that might even explain the
Standard Model of particle physics!  Everyone dropped
eleven-dimensional supergravity like a hot brick.  The mood of the
times was encapsulated by Nobel Laureate Murray Gell-Mann (inventor of
the quarks) in his closing address at the 1984 Santa Fe Meeting, when he said:
``Eleven Dimensional Supergravity (Ugh!)''.

\section{The liberty of doubt}

So ten dimensions were riding high but eleven dimensions were in the
doghouse.  Nevertheless, a small band of theorists clung to the idea
that eleven dimensions must somehow feature in the final theory.
Notwithstanding the euphoria that surrounded string theory, they allowed
themselves the {\it liberty of doubt} about string theory as it was
currently formulated and began to ask some awkward questions \cite{Duff2}:

* {\it The uniqueness problem}

Theorists love {\it uniqueness}; they like to think that the
ultimate {\it Theory of Everything} \cite{Weinberg} will one day
be singled out, not merely because all rival theories are in
disagreement with experiment, but because they are mathematically
inconsistent.  In other words, that the universe is the way it is
because it is the only possible universe.  But string theories are
far from unique.  Already in ten dimensions there are five
mathematically consistent theories as shown in Figure
\ref{Mdiagram}: the Type $I$ $SO(32)$, the heterotic $SO(32)$, the
heterotic $E_8\times E_8$, the Type $IIA$ and the Type $IIB$.
(Type $I$ is an {\it open} string in that its ends are allowed to
move freely in spacetime; the remaining four are {\it closed
strings} which form a closed loop.) If we are looking for a unique
all-embracing theory, this seems like an embarrassment of riches.

The situation becomes even worse when we consider compactifying
the extra six dimensions. There seem to be billions of different ways of
compactifying the string from ten dimensions to four (billions of
different Calabi-Yau manifolds) and hence billions of competing
predictions of the real world (which is like having no predictions at
all). This aspect of the uniqueness problem is called the {\it
vacuum-degeneracy problem}. One can associate with each different
phase of a physical system a {\it vacuum state}, so called because it
is the quantum state corresponding to no real elementary particles at
all.  However, according to quantum field theory, this vacuum is
actually buzzing with virtual particle-antiparticle pairs that are
continually being created and destroyed and consequently such vacuum
states carry energy. The more energetic vacua, however, should be unstable
and eventually decay into a (possibly unique) stable vacuum with the least
energy, and this should describe the world in which we live.
Unfortunately, all these Calabi-Yau vacua have the same energy and the
string seems to have no way of preferring one to the other.  By focusing
on the fact that strings are formulated in ten spacetime dimensions and
that they unify the forces at the Planck scale,  many critics of string
theory fail to grasp this essential point. The problem is not so much that
strings are unable to produce four-dimensional models like the Standard
Model with quarks and leptons held together by gluons, $W$-bosons, $Z$
bosons and photons and of the kind that can be tested experimentally in
current or foreseeable accelerators. On the contrary, string theorists can
dream up literally billions of them! The problem is that they have no way
of discriminating between them. What is lacking is some dynamical mechanism
that would explain why the theory singles out one particular Calabi-Yau
manifold and hence why we live in one particular vacuum; in other words,
why the world is the way it is.  Either this problem will not be solved,
in which case string theory will fall by the wayside like a hundred other
failed theories, or else it will be solved and string theory will be put
to the test experimentally. Neither string theory nor $M$-theory is
relying for its credibility on building thousand-light-year accelerators
capable of reaching the Planck energy, as some {\it End-of-Science}
Jeremiahs have suggested.

* {\it The dimension problem}

An apparently different reason for having mixed
feelings about superstrings, of course, especially for those who had
been pursuing Kaluza-Klein supergravity prior to the 1984 superstring
revolution, was the dimensionality of spacetime.  If supersymmetry
permits eleven spacetime dimensions, why should the theory of
everything stop at ten? Richard Feynman always used to say that, in
physics, whatever is not forbidden is compulsory.

* {\it The membrane problem}

\begin{figure}\centering\includegraphics[scale=1]{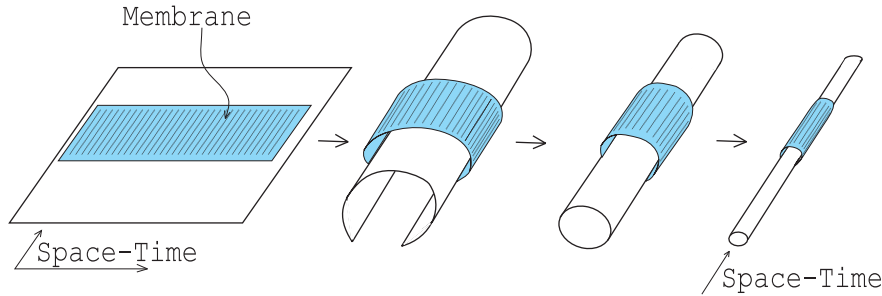}
\caption{A ten-dimensional string interpreted as
an eleven-dimensional membrane wrapped around the extra
dimension.}\label{Wrapping}\end{figure}

This problem rose to the surface again in 1987
when Eric Bergshoeff of the University of Groningen, Ergin Sezgin, now
at Texas A\&M University, and Paul Townsend from the University of
Cambridge discovered {\it The eleven-dimensional supermembrane}.  This
membrane has two spatial dimensions and moves in a spacetime dictated by
our old friend: eleven-dimensional supergravity!  It can either be a
bubble-like extended object or an infinite sheet. In the same year,
moreover, Paul Howe (King's College, London University), Takeo Inami
(Kyoto University), Kellogg Stelle (Imperial College) and I were then
able to show that if one of the eleven dimensions is a circle, then we
can wrap one of the membrane dimensions around it so that, if the
radius of the circle is sufficiently small, it looks like a string in
ten dimensions.  See Figure \ref{Wrapping}.  In fact, it yields
precisely the Type $IIA$ superstring.  This suggested to us that maybe
the eleven-dimensional theory was the more fundamental after all,
though this was not a popular view among the latter-day {\it
Flatlanders} who refused to acknowledge this extra
dimension\footnote{See Appendix.}.

\section{Supermembranes}
\label{Supermembranes}

Membrane theory has a strange history which goes back even further than
strings. The idea that the elementary particles might correspond to
modes of a vibrating membrane was put forward originally in 1960 by the
British Nobel Prize winning physicist Paul Dirac, a giant of twentieth
century science who was also responsible for two other daring postulates:
the existence of {\it anti-matter} and the existence of {\it magnetic
monopoles}. ( Anti-particles carry the same mass but opposite charge from
particles and were discovered experimentally in the 1930s. Magnetic
monopoles carry a single magnetic charge and to this day have not yet been
observed.  Interestingly enough, however, they provide an alternative
reason why electric charge should be quantized.  Once again,
\begin{equation}
e_n=ne,
\label{chargeDirac}
\end{equation}
just as Klein had said, but according to Dirac the fundamental
charge $e$ is now given by
\begin{equation}
e=2\pi/g
\label{charge2}
\end{equation}
where $g$ is the magnetic charge.  This later gave rise to
speculations that Nature might be invariant under an electric-magnetic
duality that exchanges electric particles and magnetic monopoles.  As
we shall see, both the Klein rule and the Dirac rule feature
prominently in $M$-theory, as does electric-magnetic duality.)

When string theory came along in the 1970s, there were some
attempts to revive Dirac's membrane idea but without much success. The
breakthrough did not come until 1986 when James Hughes, Jun Liu and
Joseph Polchinski of the University of Texas showed that, contrary to
the expectations of certain string theorists, it was possible to
combine the membrane idea with supersymmetry: the {\it supermembrane}
was born.

\begin{figure}\centering\includegraphics[scale=1]{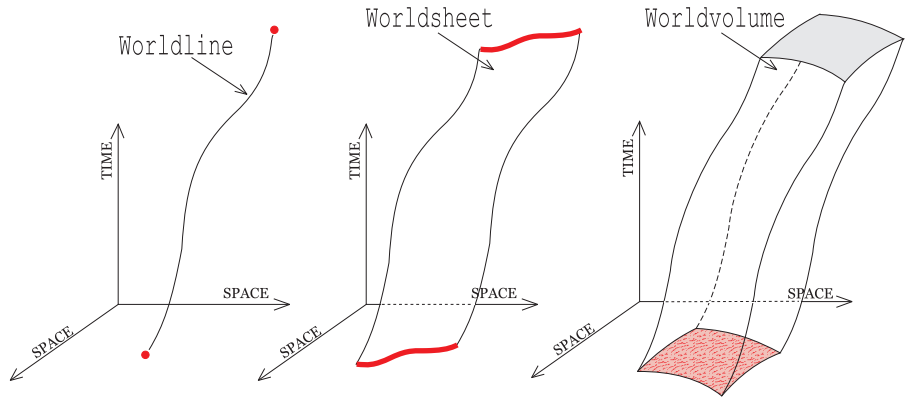}
\caption{Particles, strings and membranes.}\label{worldvolume}\end{figure}

Consequently, while all the progress in superstring theory was being made a
small but enthusiastic group of theorists were posing a seemingly very
different question: Once you have given up $0$-dimensional particles in
favor of $1$-dimensional strings, why not $2$-dimensional membranes or in
general $p$-dimensional objects (inevitably dubbed {\it $p$-branes})?
Just as a $0$-dimensional particle sweeps out a $1$-dimensional {\it
worldline} as it evolves in time, so a $1$-dimensional string sweeps out a
$2$-dimensional {\it worldsheet} and a $p$-brane sweeps out a
$d$-dimensional {\it worldvolume}, where $d=p+1$.  See
Figure \ref{worldvolume}.

\begin{figure}\centering\includegraphics[scale=0.9]{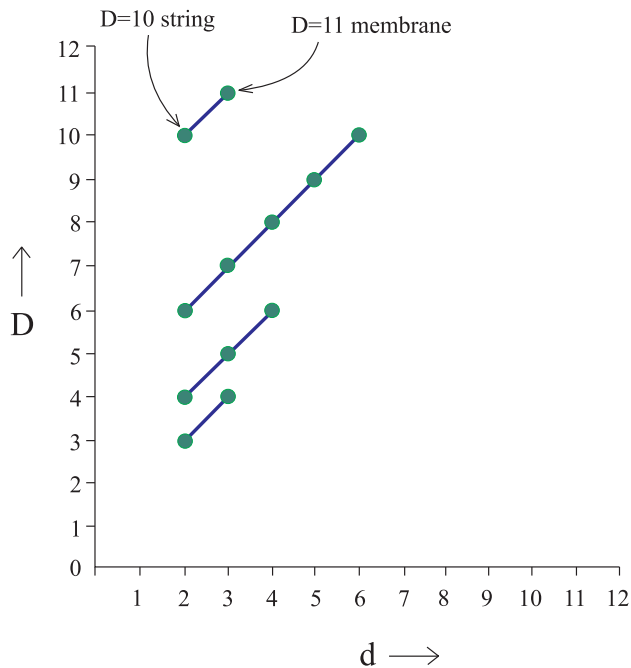}
\caption{The branescan, circa 1987.}\label{Branescan}\end{figure}

Of course, there must be enough
room for the $p$-brane to move about in spacetime, so $d$ must be less
than the number of spacetime dimensions $D$.  In fact supersymmetry
places further severe restrictions both on the dimension of the
extended object and the dimension of spacetime in which it lives.  One
can represent these as points on a graph where we plot spacetime
dimension $D$ vertically and the $p$-brane dimension $d=p+1$
horizontally.  This graph is called the {\it brane-scan}.  See Figure
\ref{Branescan}.  Curiously enough, the maximum spacetime dimension
permitted is eleven, where Bergshoeff, Sezgin and Townsend found their
$2$-brane.  In the early 80s Green and Schwarz had showed that
spacetime supersymmetry allows classical superstrings moving in
spacetime dimensions $3,4,6$ and $10$.  (Quantum considerations rule
out all but the ten-dimensional case as being truly fundamental.  Of
course some of these ten dimensions could be curled up to a very tiny
size in the way suggested by Kaluza and Klein.  Ideally six would be
compactified in this way so as to yield the four spacetime dimensions
with which we are familiar.) It was now realized, however, that there
were twelve points on the scan which fall into four sequences ending
with the superstrings or $1$-branes in $D=3,4,6$ and $10$, which were
now viewed as but special cases of this more general class of
supersymmetric extended object.

Other branes were later added to the scan.  For example, in 1991,
after Stelle and I had shown how the electric 2-brane emerges as a
solution of $D=11$ supergravity, Rahmi Guven, from University of the
Bosporus, discovered an eleven-dimensional super $5$-brane which acts
as its magnetic dual.

Notwithstanding these and subsequent results, the
supermembrane enterprise was ignored by most adherents of
conventional superstring theory\footnote{A notable exception was Paul
Townsend who continued to keep the faith and whose 1994 paper with
Christopher Hull of Queen Mary College, London, did much to convert
the sceptics.}.  Those who had worked on
eleven-dimensional supergravity and then on supermembranes spent the
early eighties arguing for {\it spacetime} dimensions greater than
four, and the late eighties and early nineties arguing for {\it
worldvolume} dimensions greater than two.  The latter struggle was by
far the more bitter\footnote{One string theorist I know would
literally cover up his ears whenever the word ``membrane'' was
mentioned within his earshot!  Indeed, I used to chide my more
conservative string theory colleagues by accusing them of being unable
to utter the M-word.}.

\section{M theory and duality}

The uniqueness problem and the dimension problem were suddenly solved
simultaneously by Witten in his, by now famous, talk at the University
of Southern California in February 1995.  Witten put forward a
convincing case that the distinction we used to draw between the five
consistent string theories was merely an artifact of our approximation
scheme and that when looked at exactly, there was really only one
theory, which subsumed all the others.  Moreover this theory was a
supersymmetric theory in eleven dimensions!  In fact, when viewed at
distances much larger than the Planck length, it is approximated by
our old friend eleven-dimensional supergravity!

Curiously enough, however, Witten still played down the importance of
supermembranes.  But it was only a matter of time before he too
succumbed to the conclusion that we weren't doing just string theory
any more!  In the coming months, literally hundreds of papers appeared
on the internet confirming that, whatever this eleven-dimensional theory
may be, it certainly involves supermembranes in an important way.
Consequently Witten dubbed it {\it M-theory} ``where M stands for
Magic, Mystery or Membrane according to taste''.

Even the chiral $E_8 \times E_8$ string, which according to Witten's
earlier theorem could never come from eleven-dimensions, was given an
eleven-dimensional explanation by Petr Horava (Princeton University) and
Witten. The no-go theorem is evaded by compactifying not on a circle (which
has no ends), but on a line-segment (which has two ends).  It is ironic
that having driven the last nail into the coffin of eleven-dimensions (and
having driven Gell-Mann to utter ``Ugh!''), Witten was the one to pull
the nail out again! He went on to argue that if the size of this
one-dimensional space is large compared to the six-dimensional space,
then our world is approximately five-dimensional.  This may have
important consequences for confronting $M$-theory with experiment.

What do we now know with M-theory that we did not know with old-fashioned
string
theory?  Here are a few examples, references to which may be found in
\cite{Duff0}.

1) Electric-magnetic duality in $D=4$ is a consequence of
membrane/fivebrane duality in $D=11$.  Moreover, the Klein
charge quantization rule and the Dirac charge quantization rule play dual
roles.  They are thus complementary rather than contradictory.

2) {\it Exact} electric-magnetic duality, first proposed for
maximally supersymmetric quark-gluon theories, has been extended
to {\it effective} duality by Nathan Seiberg (Princeton) and
Witten to less supersymmetric theories: the so-called
Seiberg-Witten theory. This has been very successful in providing
the first proofs of quark confinement (albeit in the
as-yet-unphysical super QCD) and in generating new pure
mathematics on the topology of four-manifolds. Seiberg-Witten
theory and other dualities of Seiberg may, in their turn, be
derived from M-theory.

3) Indeed, it seems likely that all supersymmetric quantum field
theories with any symmetry, and their spontaneous symmetry breaking, admit a
geometrical interpretation within M-theory as the worldvolume fields
that propagate on the common intersection of stacks of p-branes
wrapped around various holes of the compactified dimensions, with the
Higgs field given by the brane separations.

4) In string theory, the vacuum degeneracy problems arises because
there are billions of Calabi-Yau vacua which are distinct according to
classical topology.  Like higher-dimensional Swiss cheeses, each can
have different number of $p$-dimensional holes.  This results in many
different kinds of four-dimensional Standard-like Models with
different symmetries, numbers of families and different choices of quarks
and
leptons.  Moreover, M-theory introduces new effects which allow many
more possibilities, making the degeneracy problem apparently even
worse.  However, most (if not all) of these manifolds are in fact
smoothly connected in M-theory by shrinking the $p$-branes that can
wrap around the $p$-dimensional holes in the manifold and which appear
as black holes in spacetime.  As the wrapped-brane volume shrinks to
zero, the black holes become massless and effect a smooth transition
from one Calabi-Yau manifold to another.  Although this does not yet
cure the vacuum degeneracy problem, it puts it in a different light.
The question is no longer why we live in one topology rather than
another but why we live in one particular corner of the unique
topology.  This may well have a simpler explanation.
A fuller discussion may be found in Brian Greene's book \cite{Greene}.

Another interconnection was uncovered by Polchinski who realized
that the Type $II$ super $p$-branes carrying a certain kind of charge
may be identified with the so-called Dirichlet-branes (or
$D$-branes, for short) that he had studied some years ago by looking
at strings with unusual boundary conditions.  Dirichlet was a French
mathematician who first introduced such boundary conditions.  These
$D$-branes are just the surfaces on which open strings can end.  This
$D$-brane technology has opened up a whole new chapter in the history
of supermembranes.

5) Thus another by-product of these membrane breakthroughs has been an
appreciation of the role played by black holes in particle physics and
string theory. In fact they can be regarded as black branes wrapped
around the compactified dimensions. These black holes are tiny $(10^{-35}$
meters) objects; not the multi-million solar mass objects that are
gobbling up galaxies. However, the same physics applies to both and there
are strong hints that M-theory may even clear up many of the
apparent paradoxes of quantum black holes raised by Stephen Hawking in
Cambridge.  Ever since the 1970's, when Hawking used macroscopic arguments
to
predict that black holes have an entropy equal to one quarter the area
of their event horizon, a microscopic explanation has been lacking.
But treating black holes as wrapped $p$-branes, together with the
realization that Type II branes have a dual interpretation as
Dirichlet branes, allowed Andrew Strominger and Cumrun Vafa of Harvard
to make the first microscopic prediction in complete
agreement with Hawking.  The fact that M-theory is clearing up some
long standing problems in quantum gravity gives us confidence that we
are on the right track.

6) It is known that the strengths of the four forces change with
distance.  In supersymmetric extensions of the Standard Model, one
finds that the strengths of the strong, weak and electromagnetic
forces all meet at about at very short distances entirely
consistent with the idea of Grand Unification.  The strength of the
gravitational force also almost meets the other three, but not quite.
This near miss has been a source of great interest, but also
frustration.  However, in a universe of the kind envisioned by Witten,
spacetime is approximately a narrow five dimensional layer bounded by
four-dimensional walls\footnote{If this picture is correct, we really
need only five dimensions with the other six going along for the ride.
Why should Nature behave like this?  The only good answer to this
question I could find is in Mother Goose's Nursery Rhymes:

{\it Nature requires five,

Custom allows seven,

Idleness takes Nine,

And Wickedness Eleven.}}.  The particles of the Standard Model live on
the walls but gravity lives in the five-dimensional bulk.  As a
result, it is possible to choose the size of this fifth dimension so
that all four forces meet at this common scale.  Note that this is
much larger than the Planck length, so gravitational effects may be
much closer than we previously thought; a result that would have all
kinds of cosmological consequences.  Indeed, this has lead to a revival of a
variation on the Kaluza-Klein theme whereby our universe is a
$3$-brane embedded in a higher-dimensional spacetime.  Thus the strong,
weak and electromagnetic forces might be confined to the worldvolume
of the brane while gravity propagates in the bulk.  It has recently
been suggested that, in such schemes, the extra dimension might be
much larger than $10^{-35}~{\rm meters}$ and may even be a large as a
millimeter \cite{Dimopoulos}.  In yet another variation due to Lisa
Randall of Harvard and Raman Sundrum of Johns Hopkins, the extra fifth
dimension is infinite.

Thus branes are no longer the ugly-ducklings of string theory, but now
play center stage in theoretical physics: as the microscopic
constituents of M-theory, as the higher-dimensional progenitors of
black holes and as entire universes in their own right.

\section{So what is $M$-theory?}

{\it So are we quarks, strings, branes or what?}

\medskip
New York Times, Tuesday September 22, 1998
\bigskip

I hope to have convinced you that the answer involves ``branes or what'',
but there is still no final answer to the exact definition of M-theory,
although several different proposals have been made.  Is M-theory to
be regarded literally as membrane theory?  In other words should we
attempt to ``quantize'' the eleven dimensional membrane in some, as
yet unknown way?  Personally, I think the jury is still out on whether
this is the right thing to do.

Tom Banks and Stephen Shenker at Rutgers together with Willy Fischler
from the University of Texas and Lenny Susskind at Stanford have even
proposed a rigorous definition of $M$-theory known as M(atrix) theory
which is based on an infinite number of Dirichlet $0$-branes.  In this
picture spacetime is a fuzzy concept in which the spacetime
coordinates $x,y,z,...$ are matrices that do not commute e.g.  $xy
\neq yx.$ This approach has generated great excitement but does yet
seem to be the last word.  It works well in high dimensions but as we
descend in dimension it seems to break down before we reach the real
four-dimensional world.

Another interesting development has been provided by Juan Maldacena
at Princeton, who has suggested that $M$-theory compactified to a particular
$D$-dimensional spacetime, including all its gravitational
interactions, may be completely described by a non-gravitational
theory that resides on the $(D-1)$-dimensional boundary of that
spacetime.  This holds promise not only of a deeper understanding of
M-theory, but may also throw light on aspects of the theories that
live on the boundary, which in some circumstances can include the
kinds of 4-dimensional quark theories that govern the strong nuclear
interactions such as QCD.  Many theorists are understandably excited
about this correspondence because of what it can teach us about QCD.
In my opinion, however, this is, in a sense, a diversion from the
really fundamental question: what is M-theory?  So my hope is that
Maldacena's idea will be a two-way process and that it will also teach
us more about M-theory.

$M$-theory has sometimes been called the {\it Second Superstring
Revolution}, but I feel this is really a misnomer.  It certainly
involves new ideas every bit as significant as those of the 1984
string revolution, but its reliance upon supermembranes and eleven
dimensions makes it is sufficiently different from traditional string
theory to warrant its own name.  One cannot deny the tremendous
historical influence of superstrings on our current perspectives.
Indeed, it is the pillar upon which our belief in a quantum consistent
M-theory rests.  In my opinion, however, the focus on one-dimensional
objects moving in a ten-dimensional spacetime, that prevailed during
the period 1984-1995, will ultimately be seen to be a small corner of $M$-theory
\cite{Duff}.  Indeed, future historians may well judge this era as a time
when theorists were like boys playing by the sea shore, and diverting
themselves with the smoother pebbles or prettier shells of
ten-dimensional superstrings while the great ocean of
eleven-dimensional $M$-theory lay all undiscovered before them.

\section{Acknowledgments}

\noindent

I would like to thank Dean Shirley Neuman and the University of
Michigan Administration for conferring on me the title of Oskar Klein
Collegiate Professor of Physics.  It is indeed a great honor.  I
would also like to thank the Physics Department Chair, Ctirad Uher for
recruiting me to Michigan in 1999; without him I would not be here.
What impressed me most about the University was the intellectual
caliber of the faculty and I would also like to acknowledge my colleagues in
the Physics Department for providing such a stimulating environment in
which to conduct research.  Thanks to Stanley Deser and Jim Liu for
suggesting improvements in the paper and to Alex Batrachenko for
invaluable help with the pictures and diagrams.
Living with a physicist is not easy.  So (and at the risk of this sounding
more like the Oscars than the Oskars), I would also like to
thank my wife Lesley, my daughter Jessica and my son Matthew, for
putting up with me.

\newpage
\appendix

\section{Flatland, Modulo 8}

\subsection{Modern Translation}

Edwin A.  Abbott (1838-1926) was a English clergyman and
Shakespearean scholar who wrote his satire, {\it Flatland: A Romance of Many
Dimensions} \cite{Abbott}, in 1884.  Hence it is first necessary to provide
a dictionary of updated terminology:
\bigskip

2-dimensional Flatland\ldots\ldots 10-dimensional String Theory

Line-like objects\ldots\ldots Strings

A.  Square\ldots\ldots A Humble String Theorist, Hero of Our Story

His sons, the Pentagons\ldots\ldots His Postdocs

His grandsons, the Hexagons\ldots\ldots His Graduate Students

Polygons\ldots\ldots Full Professors

Circles\ldots\ldots The Sultans of String

High Council\ldots Recipients of the MacArthur ``Genius'' Awards in
String Theory

President of the Council\ldots\ldots Edgar Whittington, Fields Medalist

The Secret Archives\ldots\ldots arXiv.org

The Prefect\ldots\ldots Jaguar Quark-Mann, Nobel Laureate

The Palace of the Prefect\ldots\ldots California Institute of Technology

3-dimensional Spaceland\ldots\ldots 11-dimensional M-theory

Space-like objects\ldots\ldots Membranes

Sphere\ldots\ldots An M-theorist

Brightness\ldots\ldots Electric Charge, or Eleventh Component of
Momentum

Height\ldots\ldots Radius of the Eleventh Dimension

Policemen\ldots\ldots Journalists

Land of 4-dimensions\footnote{A speculative idea, whose status is
still obscure at the end of the story.}\ldots\ldots 12-dimensional
F-theory

Point-like objects\ldots\ldots Particles

King of Pointland\ldots\ldots Sheldonian Glasgow, Nobel Laureate

The Abyss of No Dimensions\ldots\ldots Quantum Field Theory

\bigskip

The story has been shortened but, apart from the above updates, the words
below are exactly as Abbott wrote them.  (In the first part of the
book, he lampoons the attitude to women in Victorian society,
especially in Science.  Readers are invited to update that part.)
For dramatic effect, Abbott sets the scene on New Year's Eve, 1999.
Although this was remarkably prescient, historians of science now
believe that these events actually took place ten years earlier.  Now
read on.

\subsection{Concerning a stranger from the Eleventh Dimension}

 It was the last day of the 1999th year of our era.  The pattering
of the rain had long ago announced nightfall; and I was sitting in the
company of my wife, musing on the events of the past and the prospects
of the coming year, the coming century, the coming Millennium.

I was rapt in thought, pondering in my mind some words that had casually
issued from the mouth of my youngest graduate student, a most
promising young man of unusual brilliancy.  I had been induced to reward
him by giving him a few hints on Arithmetic, as applied to String Theory.

``I suppose an Eleventh Dimension must mean something in String Theory;
what does it mean?''  ``Nothing at all,'' replied I, ``not at least in
String Theory; for String Theory has only Ten Dimensions.'' ``Go,'' said I,
a little ruffled by this interruption; ``if you
would talk less nonsense, you would remember more sense.''

So my
graduate student disappeared in disgrace; and I sat by my wife's
side, endeavouring to form a retrospect of the year 1999 and of the
possibilities of the year 2000, but not quite able to shake off the
thoughts suggested by the prattle of my bright little graduate student.
Only a few sands now remained in the half-hour glass.

Straightway, I became conscious of a Presence in the room, and a
chilling breath thrilled through my very being.

``The boy is a fool, I say; Eleven Dimensions can have no meaning in
String Theory.'' At once there came a distinctly audible reply,``The
boy is not a fool; and Eleven Dimensions has an obvious Geometrical
meaning.'' What was my horror when I saw before me a Figure!  I should
have thought it a String Theorist only that it seemed to change its
size in a manner impossible for a String Theorist or for any regular
Figure of which I had had experience.

``I am, in a certain sense a String Theorist,'' replied the Voice,
``and a more perfect String Theorist than any in Ten Dimensions; but
to speak more accurately, I am an M-theorist.''

I glanced at the half-hour glass. The last sands had fallen.  The third
Millennium had begun.

\subsection{How the stranger vainly endeavoured to reveal to me in
words the mysteries of Eleven Dimensional M-theory}

 {\it I}: Would your Lordship indicate or explain to me in what direction
is the Eleventh Dimension, unknown to me.

 {\it Stranger}: I came from it.  It is up above and down below.

{\it  I}: Your Lordship jests.

{\it Stranger}: I am in no jesting humour.  I tell you that I come from
M-theory, or, since you will not understand what M-theory means, from
the Land of Membranes whence I lately looked down upon your Plane that
you call String forsooth.  From that position of advantage I discerned
all that you speak of as a Membrane.

{\it I}: Such assertions are easily made, my Lord.

 {\it Stranger}: But not easily proved, you mean. But I mean to prove mine.

 {\it I}: But am I to suppose that your Lordship gives to Electric Charge
the title of Dimension, and that what we call ``Charge'' you call
eleventh component of momentum?

{\it Stranger}: No, indeed.  By the eleventh coordinate,
I mean a dimension like your length: only, with you, the radius of the
eleventh dimension is not so easily perceptible, being extremely small.  For
even an
M-theorist-which is my proper name in my own country- if he manifest
himself at all to an inhabitant of Ten Dimensions-must needs manifest
himself as a String Theorist.

Every reader in M-theory will easily understand that my mysterious
Guest was speaking the language of truth and even simplicity.  But to
me, proficient though I was in String Mathematics, it was by no means a
simple matter.

``Monster,'' I shrieked, ``be thou juggler, enchanter, dream or devil,
no more will I endure thy mockeries. Either thou or I must perish.''
And saying these words I precipitated myself upon him.

\subsection{ How the M-theorist, having in vain tried words,
resorted to deeds}

 {\it M-theorist}: Why do you refuse to listen to reason?  I had
hoped to find in you-as being a man of sense and an accomplished
mathematician- a fit apostle for the Gospel of the Eleven Dimensions,
which I am allowed to preach once only in a thousand years: but now I
know not how to convince you.  Away from me, or you must go with me-
whither you know not-into the Land of Eleven Dimensions!

``Fool! Madman!'' I exclaimed.

``Ha! Is it come to this?'' thundered the Stranger: ``then meet your
fate: out of your Plane you go. Once, twice, thrice! `Tis done!''

\subsection{How I came to the Eleventh Dimension and what I saw}

 An unspeakable horror seized me.  There was a darkness; then a dizzy
 sickening sensation of sight that was not like seeing.  When I could find
 voice, I shrieked aloud in agony, ``Either this is madness or it is
 Hell.'' It is neither replied the voice of the M-theorist, ``it is
 Knowledge; it is Eleven Dimensions: open your eyes once again and try
 to look steadily.''

I looked, and behold, a new world! Something-for which I had no
words; but you, my readers in M-theory, would call a Membrane.

Bewildered though I was by my Teacher's enigmatic utterance, I no
longer chafed against it, but worshipped him in silent adoration. He
continued with more mildness in his voice. ``Distress not yourself if
you cannot at first understand the deep mysteries of Eleven Dimensions.
By degrees they will dawn upon you. But enough of this. Look yonder.
Do you know that building?''

I looked, and afar off I saw an immense Polygonal structure, in which
I recognised the General Assembly Hall of the States of String Theory;
and I perceived that I was approaching the great Metropolis.

It was now morning, the first hour of the first day of the two
thousandth year of our era.  Acting, as was their wont, in strict accordance
with precedent, the highest String Theorists of the realm were meeting in
solemn conclave\footnote{Abbott can only be referring to the {\it Strings
2000}
conference, but historians favor {\it Strings 1990}, whose Organizing
Committee is known to have passed the following resolution (with one
dissenting
vote): ``The word {\it Membrane} shall nowhere appear in the
topics listed on the {\it Strings 1990} conference poster.''}.

The minutes of the previous meeting were now read: ``Whereas the States
had been troubled by divers ill-intentioned
persons pretending to have received revelations from another World,
and professing to produce demonstrations whereby they had instigated
to frenzy both themselves and others, it had been for this cause
unanimously resolved by the MacArthur Fellows that on the first day of each
millenary, special injunctions be sent to the Prefects in several
districts of String Theory, to make strict search for such misguided
persons and without formality of mathematical examination, to destroy
all such..''

``You hear your fate'', said the M-theorist to me, while the MacArthur
Fellows were passing for the third time the formal resolution.  `` Death
or imprisonment awaits the Apostle of the Eleven Dimensions''.  ``Not
so, replied I, ``the matter is now so clear to me, the nature of real
space so palpable, that methinks I could make a child understand it.
Permit me but to descend at this moment and enlighten them.'' ``Not
yet'', said my Guide, `` the time will come for that.  Meantime I must
perform my mission.  Stay thou there in thy place.'' Saying these
words, he leaped with great dexterity into the sea (if I may so call
it) of String Theory, right in the midst of the ring of MacArthur
Fellows.  ``I come,'' cried he,  ``to proclaim that there is a land of
Eleven Dimensions.''

I could see many of the younger Fellows start back in manifest
horror.

``My Lords,'' said President Edgar Whittington to the Junior String
Theorists,
``there is not the slightest need for surprise; the secret archives,
to which I alone have access, tell me that a similar occurrence
happened on the last two millennial commencements\footnote{Some
historians place these secret archives at Los Alamos National
Laboratory, but they now seem to have disappeared.}.  You will, of
course, say nothing of these trifles outside the Cabinet.''

Raising his voice, he now summoned the guards. ``Arrest the
journalists; gag them.  You know your duty.'' After he had consigned
to their fate the wretched journalists-ill-fated and unwilling
witnesses of a State-secret which they were not permitted to reveal-he
again addressed the String Theorists.  `` My Lords, the business of
the Fellows being concluded, I have only to wish you happy New Year.''

\subsection{How, though the M-theorist shewed me other mysteries of
Eleven Dimensions, I still desired more; and what came of it}

 Once more we ascended into Eleven Dimensions.  ``Hitherto,'' said
the M-theorist, ``I have shewn you naught save Branes.  Now I must
introduce you to Stacks, and reveal to you the plan upon which they
are constructed.  Behold this multitude of square cards.  See, I put
one upon the other, not as you supposed, Northward of the other, but
{\it on} the other.  Now a second, now a third.  See, I am building up
a Stack by a multitude of Branes parallel to one another.  Now the
Stack is complete.

Were I to give the M-theorist's explanation of these matters, succinct
and clear though it was, it would be tedious to an inhabitant of
Eleven Dimensions, who knows these things already.  Suffice it that I
could now readily distinguish between a String and a Membrane.

This was the Climax, the Paradise, of my strange eventful history.
Henceforth I have to relate the story of my miserable Fall: most
miserable, yet surely most undeserved! For why should the thirst for
knowledge be aroused, only to be disappointed and punished? My
volition shrinks from the painful task of recalling my humiliations;
yet, like a second Prometheus, I will endure this and worse, if by
any means I may arouse in the interiors of String Humanity a
spirit of rebellion against the Conceit which would limit our
Dimensions to Ten or even Eleven.

{\it I}: My Lord, your own wisdom has taught me to aspire to One even more
great, more beautiful, and more closely approximate to Perfection than
yourself. As you yourself, superior to all Ten Dimensional forms,
combine many strings in One, so doubtless there is One above you who
combines many Membranes in One Supreme Existence, surpassing even
M-theory. And even as we, who are now in Eleven Dimensions, look down
on Ten Dimensions, so of a certainty there is yet above us some
higher, purer F-theory - some yet more spacious Space, some more
dimensionable Dimensionality.

{\it M-theorist}: Pooh! Stuff! Enough of this trifling!  Time is short, and
much
remains to be done before you are fit to proclaim the Gospel of Eleven
Dimensions to your blind benighted countrymen in String Theory.

{\it I}: What therefore more easy than now to take his servant on a second
journey into the blessed region of the Twelfth Dimension?

{\it M-theorist}: But where is this land of Twelve Dimensions?

{\it I}: I know not: but doubtless my Teacher knows.

{\it M-theorist}: Not I. There is no such land. The very idea of it is
utterly inconceivable.

My words were cut short. Down! Down! Down! I was rapidly descending;
and I knew that a return to String Theory was my doom.

\subsection{How the M-theorist encouraged me in a vision}

When I was at last by myself, a drowsy sensation fell on me; but
before my eyes closed I endeavoured to reproduce the Eleventh
Dimension.  During my slumber I had a dream.

``Look yonder,'' said my guide, ``in String Theory thou hast lived;
thou hast soared with me to the heights of M-theory; now, in order to
complete the range of thy experience, I conduct thee downward to the
lowest depth of existence, even to Quantum Field Theory, the Abyss of
No Dimensions.

``Behold yon miserable creature.  That Particle Physicist,
Sheldonian Glasgow, is a Being like ourselves, but confined to the
non-dimensional gulf.  He is himself his own World, his own Universe;
of any other than himself he can form no conception; he knows neither String
nor Membrane, for he has no experience of them.  Yet mark his perfect
self-contentment.

``Can you not startle the little thing out of its complacency?'' said
I.  ``Tell it what it really is, as you told me; reveal to it the
narrow limitations of Pointland, and lead it up to something higher.''
``That is no easy task,'' said my master.  ``Let us leave this King of
Pointland to the ignorant fruition of his omnipresence and
omniscience: nothing that you or I can do can rescue him from his
self-satisfaction.''

After this, as we floated gently back to Ten Dimensional String Theory, I
could hear the mild voice of my Companion pointing the moral of my
vision, and stimulating me to aspire, and to teach others to aspire.

 \subsection{How I tried to teach the theory of Eleven
Dimensions to my graduate student, and with what success}

I awoke rejoicing, and began to reflect on the glorious career
before me.  I would go forth, methought, at once, and evangelize the
whole of String Theory.

Just as I had decided on the plan of my operations, I heard a loud
voice.  It was the a herald's proclamation.  Listening attentively, I
recognized the words of the Resolution of the MacArthur Fellows,
enjoining the arrest, imprisonment, or execution of any one who
should pervert the minds of the people by delusions, and by professing
to have received revelations from another World.

I reflected.  This danger was not to be trifled with.  It would be better
to avoid it by omitting all mention of my Revelation.

My Postdocs were men of character and standing, and physicists of no
mean reputation, but not great in mathematics, and, in that respect,
unfit for my purpose.  But it occurred to me that my young docile
graduate student, with his mathematical turn, would be a most suitable
pupil.  Discussing the matter with him, a mere boy, I should be in
perfect safety; for he would know nothing of the proclamation of the
MacArthur Fellows; whereas I could not feel sure that my Postdocs- so
greatly did their patriotism and reverence for the Sultans of String
predominate over mere blind affection-might not feel compelled to hand
me over to the Prefect, Jaguar Quark-Mann, if they found me seriously
maintaining the seditious heresy of the Eleventh Dimension.

At this moment we heard once more the herald's ``Oh Yes!  Oh Yes!''
outside in the street proclaiming the resolution of the MacArthur
Fellows.  Young though he was, my graduate student-who was unusually
intelligent for his age, and bred up in perfect reverence for the
Sultans of String-took in the situation with an acuteness for which I
was quite unprepared.  He remained silent till the last words of the
proclamation had died away, and then, bursting into tears, ``Dear
Square,'' he said, ``that talking of Eleven Dimensions was only my
fun, and of course I meant nothing at all by it; and we did not know
anything then about the new Law; and I don't think I said anything about
M-theory.  How silly it is!  Ha!  Ha!  Ha!''

Thus ended my first attempt to convert a pupil to the Gospel of Eleven
Dimensions.

\subsection{How I then tried to diffuse the theory of Eleven
Dimensions by other means, and of the result}

So I devoted several months in privacy to the composition of a
treatise on the mysteries of Eleven Dimensions.  Only, with a view
of evading the Law, if possible, I spoke not of a physical
Dimension, but of Thoughtland.

Meanwhile my life was under a cloud.  All pleasures palled upon me;
all sights tantalized and tempted me to outspoken treason, because I
could not but compare what I saw in Ten Dimensions with what it really
was if seen in Eleven, and could hardly refrain from making my
comparisons aloud.

I felt that I would have been willing to sacrifice my life for the
Cause, if thereby I could have produced conviction.  But if I could not
convince my graduate student, how could I convince the highest and
most developed String Theorists in the land?

And yet at times my spirit was too strong for me, and I gave vent to
dangerous utterances.  Already I was considered heterodox if not
treasonable, and I was keenly alive to the danger of my position;
nevertheless I could not at times refrain from bursting out into
suspicious or half-seditious utterances, even among the highest Full
Professors and most developed Sultans of String society.

 At last, to complete a series of minor indiscretions, at a meeting of
our Local Speculative Society held at the Palace of the Prefect,
California Institute of Technology,- some extremely silly person
having read an elaborate paper exhibiting the precise reasons why
Providence has limited the number of Dimensions to Ten-I so far forgot
myself as to give an exact account of the whole of my voyage with the
M-theorist into Eleven Dimensions, and to the Assembly Hall in our
Metropolis, and then to Eleven Dimensions again, and of my return
home, and of everything that I had seen and heard in fact or vision.
At first, indeed, I pretended that I was describing the imaginary
experiences of a fictitious person; but my enthusiasm soon forced me
to throw off all disguise, and finally, in a fervent peroration, I
exhorted all my hearers to divest themselves of prejudice and to
become believers in the Eleventh Dimension.

Need I say that I was at once arrested and taken before the MacArthur
Fellows?

After I had concluded my defence, President Whittington, perhaps perceiving
that some of the Junior Fellows had been moved by my evident earnestness,
asked me two questions:-

1.  Whether I could indicate the direction which I meant when I used
the word {\it Eleven}?

2.  Whether I could by any diagrams indicate the figure I was pleased to
call a Membrane?

I declared that I could say nothing more, and that I must commit
myself to the Truth, whose cause would surely prevail in the end.

The President replied that I must be sentenced to perpetual
imprisonment; but if the Truth intended that I should emerge from
prison and evangelize the world, the Truth might be trusted to bring
that result to pass.

Here I am, absolutely destitute of converts, and, for aught that I can
see, the millennial Revelation has been made to me for nothing.
Prometheus was bound for bringing down fire for mortals, but I-poor
10-dimensional String Theory Prometheus-lie here in prison for bringing
down
nothing to my countrymen.  Yet I exist in the hope that these memoirs,
in some manner, I know not how, may find their way to the minds of
humanity in Some Dimension, and may stir up a race of rebels who shall
refuse to be confined to limited Dimensionality.

\subsection{Postscript}

Although Abbott never wrote about it, the story has a happy ending.
Five years later, the President of the Council made an historic
announcement at the University of Southern California lifting the ban
on Eleven Dimensions.  Fearing civil unrest, however, he continued to
proscribe discussion of Membranes.  After the String Theorists were
given time to get over the shock of Eleven Dimensions, though, even
Membranes
ceased to be taboo.  The gag on the journalists was removed, and our hero
the Square was released from prison.  Indeed, he has since been
promoted to a Polygon.

\end{document}